\documentclass[twocolumn]{aastex61}
\usepackage[utf8]{inputenc}
\usepackage{graphicx}
\usepackage{natbib}
\DeclareUnicodeCharacter{2212}{-}
\shorttitle{[CII] in the Perseus molecular cloud}
\shortauthors{Hall et al.}

\begin{document}
\setcitestyle{authoryear}

\author[0000-0002-0786-7307]{K. P. Hall}
\affiliation{Department of Astronomy, University of Wisconsin - Madison, 475 North Charter Street, Madison, WI, 53706-15821, USA}
\email{khall@astro.wisc.edu}
\nocollaboration

\author{Sne\v{z}ana Stanimirovi{\'c}}
\affiliation{Department of Astronomy, University of Wisconsin - Madison, 475 North Charter Street, Madison, WI, 53706-15821, USA}
\nocollaboration

\author{Min-Young Lee}
\affiliation{Korea Astronomy and Space Science Institute, 776 Daedeokdae-ro, 34055 Daejeon, Republic of Korea}
\affiliation{Max-Planck Institute for Radio Astronomy, Auf dem H{\"u}gel 69, 53121 Bonn, Germany}
\nocollaboration

\author{Mark Wolfire}
\affiliation{Department of Astronomy, University of Maryland, College Park, MD 20742, 0000-0003-0030-9510}
\nocollaboration

\author{Paul Goldsmith}
\affiliation{Jet Propulsion Laboratory, California Institute of Technology, 4800 Oak Grove Drive, Pasadena, CA 91109-8099, USA 0000-0002-6622-8396}
\nocollaboration

\title{{\it Herschel} 158$ \mu$\MakeLowercase{m} [CII] Observations of ``CO-dark" Gas in the Perseus Giant Molecular Cloud}




\begin{abstract}
We present observations of velocity-resolved [CII] 158 $\mu$m emission from both a dense and a more diffuse photodissociation region (PDR) in the Perseus giant molecular cloud (GMC) using the Heterodyne Instrument for the Far-Infrared onboard the {\it Herschel Space Telescope}.
We detect [CII] emission from 80\% of the total positions, with a 95\% detection rate from the dense boundary region.
The integrated intensity of the [CII] emission remains relatively constant 
across each boundary, despite the observed range in optical extinction between 1 mag and 10 mag.
This flat profile indicates a constant heating and cooling rate within both regions observed.
The integrated intensity of [CII] emission is reasonably well correlated with the neutral hydrogen (HI) column density, as well as total gas column density. This, in addition to the 80$\arcmin$ (7 pc) extent of the [CII] emission from cloud center, suggests that the HI envelope plays a dominant role in explaining the [CII] emission 
emanating from Perseus. 
We compare the [CII] and $^{12}$CO integrated intensities with predictions from a 1-D, two-sided slab PDR model and show that a simple core $+$ envelope, equilibrium model without an additional ``CO-dark'' H$_2$ component can reproduce observations well.
Additional observations are needed to disentangle how much of the [CII] emission is associated with the ``CO-dark'' H$_2$ gas, as well as constrain spatial variations of the dust-to-gas ratio across Perseus. 


\end{abstract}

\section{Introduction}
To understand star formation we need to understand the formation of giant molecular clouds (GMCs).
These future stellar nurseries are marked by boundaries, which are defined by the transition between
primarily atomic and primarily molecular gas. 
The ionization, chemistry, and heating
within these boundary regions are dominated by far-ultraviolet radiation (FUV), and have hence 
earned the name photo-dominated regions or photo-dissociation regions (PDR; \citealt{1999RvMP...71..173H}).

Within PDRs, molecular abundances are not uniformly distributed and the 
fractional abundances, such as HI/H$_{2}$, CH/H$_2$, or C$^{+}$/CO, 
vary appreciably among GMCs and even within a given GMC. 
The cause of these abundance variations are environmental effects including the interstellar
radiation field (ISRF), the cosmic ray ionization rate, density fluctuations, and
interstellar turbulence. 
Observations of PDRs with varied environments are essential to tease out 
the complex dependency of molecule formation on these
various environmental effects.
In particular, low-excitation PDRs with incident ISRFs ranging from less than to a few times the Habing field (G$_0$;
\citealt{1968BAN....19..421H}\footnote{The Habing field is a unit of energy density within a specific UV-wavelength range. 
This value, G$_0$ is equal to 5.26$\times$10$^{-14}$erg cm$^{-3}$ and is estimated over the energy range of 6-13.6 eV.}) have not been studied in as much detail as brighter PDRs with incident ISRFs greater than 10G$_0$ (e.g. Orion Bar and NGC 7023 NW, \citealt{2018A&A...615A.129J}, and Orion Molecular Cloud 1 (OMC1) \citealt{2015ApJ...812...75G}).

Molecular hydrogen (H$_2$), the most abundant molecule in the interstellar medium (ISM) (e.g. \citealt{2004ApJ...604..222C}, \citealt{2019snell_fundamentals}), does not have a permanent dipole moment and can only radiate through rotation-vibration, pure rotational quadrupole, or collision-induced dipole radiation \citep{1966ARA&A...4..207F}.
These transitions are typically weak in molecular clouds, especially 
in areas with no active star-formation. Therefore, alternative tracers have been
employed to infer the abundance and distribution of H$_2$ in GMCs where the typical kinetic gas temperature is $10-60$ K (e.g. \citealt{1997ApJ...483..210W}).

One of the most common methods of deriving H$_2$ column densities is through observing the $^{12}$CO (typically J=1-0) intensity ($I_{\rm CO}$) and scaling by a conversion factor. This conversion factor between $I_{\rm CO}$ and 
H$_2$ column density ($N_{H_2}$), is the X$_{\rm CO}$ factor. 
The typical assumed Milky Way X$_{\rm CO}$ value is 2-4$\times 10^{20}$ cm$^{-2}$/(K km s$^{-1}$) \citep{2013Bolatto}.
However, many uncertainties exist because
this method often assumes that $^{12}$CO is cospatial and interspersed evenly with H$_2$. 
Both theoretical and observational studies show that 
H$_2$ is more extended spatially than $^{12}$CO, with a larger spatial disparity at low metallicity (\citealt{2010Wolfire}, \citealt{2007Leroy}, \citealt{2014ApJ...784...80L}).
This H$_2$ gas without corresponding detectable $^{12}$CO emission is referred to as ``CO-dark" molecular gas.


The existence of ``CO-dark'' molecular
gas
has been known for over three decades \citep{1988LNP...315..168V, 1988ApJ...326L..69L}.
More recently it has been discovered that H$_2$ is not only generally more extended than CO (\citealt{2005Sci...307.1292G}; \citealt{2012ApJ...748...75L}), but  X$_{\rm CO}$ may vary appreciably across individual interstellar environments (e.g., \citealt{2011GloverMacLow}; \citealt{2011Shettya, 2011Shettyb}; \citealt{2014ApJ...784...80L}; \citealt{2018MNRAS.474.4672L}). 
There are many parameters that can cause spatial variations of X$_{\rm CO}$,
such as metallicity, the strength of the FUV ISRF, 
the internal density distribution, the total mass of the cloud, etc.
\citep{2006MNRAS.371.1865B,2010Wolfire,2011Shettya, 2011Shettyb}.
Therefore, calibrating alternative methods for constraining the H$_2$ column 
density is highly important.

 
 In a classical PDR scenario, ionized carbon exists in the atomic outer layer of a GMC which is irradiated by the ambient ISRF of the Galaxy or by the intense radiation field of a nearby OB cluster. 
Inside the surface layers of a cloud, as measured by visual extinction (A$_V$) with an A$_V\sim0.5$  (depending on gas density and the strength of the incident FUV radiation field), H$_2$ forms, but within this layer carbon still primarily exists as C$^+$.
Due to its lower abundance and less efficient self-shielding against FUV radiation in comparison to H$_2$, CO starts to form even deeper within a GMC and will be bright and abundant at A$_V>1$\footnote{We consider here the CO J = 1-0 transition at T$\leq$50 K and ISRF$<2$G$_0$, \cite{2012MNRAS.424.2599C}.}.
It is therefore expected that the 1900.5369 GHz (or 158 $\mu$m) [CII] line of ionized carbon should be a good tracer of the H$_2$ gas that has formed within the outer layers of a GMC where the abundance of $^{12}$CO is very low.

There have been many studies, several are summarized in Section \ref{sec:Background_2}, utilizing the  158 $\mu$m transition as a way of estimating the ``CO-dark" H$_2$.
Since the [CII] line can be excited in various ISM phases by collisions with multiple partners including electrons, H$^0$, and H$_2$, estimates of H$_2$ require 
high spatial and velocity resolution of multiple gas phases to disentangle the fraction of the [CII] intensity that corresponds to molecular gas.
In addition to the 158 $\mu$m [CII] line, the [CI] hyperfine transitions at 492 GHz and 809 GHz are also considered as potentially good tracers of the ``CO-dark'' H$_2$ gas, however observational data determining whether these transitions arise from diffuse or dense molecular gas are currently unclear \citep{2014A&A...571A..53B}.

\begin{figure*}
	\includegraphics[width=\textwidth]{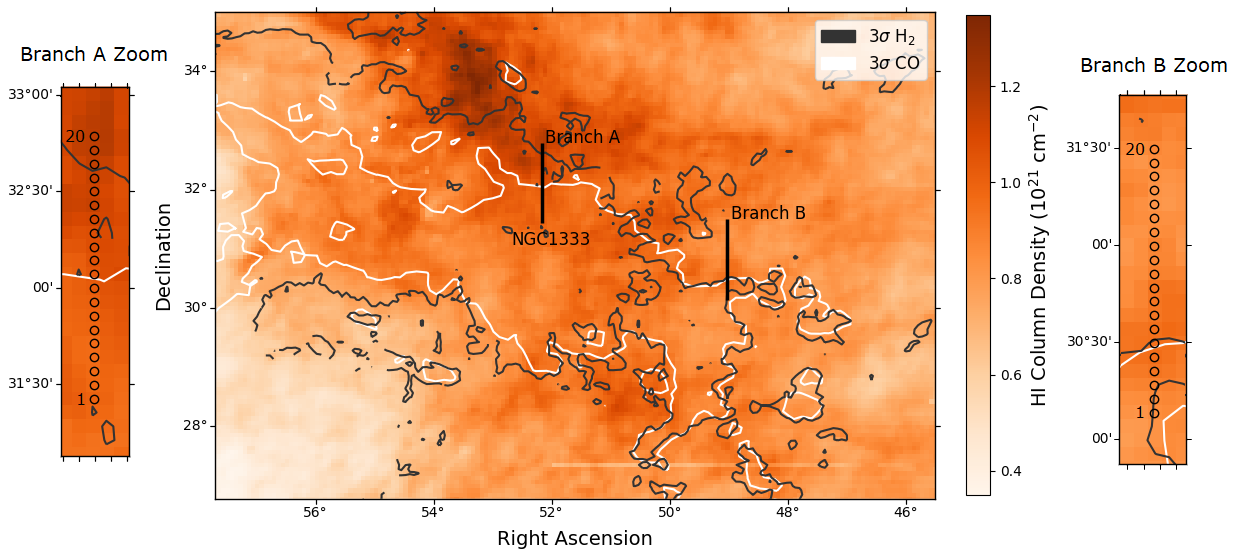}
	\caption{The HI column density image of Perseus derived from the GALFA HI Survey \citep{2011ApJS..194...20P}. Regions where we obtained [CII] spectra with \textit{Herschel} are outlined as straight black lines in the center image, while individual pointings are shown as circles on the two side-zoom-in plots. On the center image the branches are marked as ``Branch A'' and ``Branch B'', pointings A1 and B1 represent the positions of the base of the branches, the bottom most circles on the zoom-in maps, all positions subsequently are referred to by A1, A2, A3... A20
	and B1, B2, B3... B20, respectively. 
	The resolution of the [CII] observations is smaller than the displayed circles (for an accurately depicted aperture refer to Figure~\ref{f:diff_res}).
	The dark gray contour corresponds to a SNR cutoff of 3$\sigma$ for the H$_2$ surface density derived by \cite{2014ApJ...784...80L}. The white contour corresponds to a SNR cutoff of 3 for the integrated $^{12}$CO (J$=$1-0) intensity from \citealt{2001Dame} and \citealt{2006Ridge}.}
	\label{f:Map}
\end{figure*}

Another commonly used method for constraining the fraction of the ``CO-dark" H$_2$ gas is based on infrared observations (e.g. \citealt{1997A&A...328..471I}, \citealt{2001Dame}).
For example, \citeauthor{2012ApJ...748...75L} (\citeyear{2012ApJ...748...75L}; hereafter referred to as L12) combined infrared observations from IRAS 
with the GALFA-HI observations from the Arecibo radio telescope to 
estimate the distribution of H$_{2}$ across the Perseus molecular cloud  (Figure 1) under the assumption of
a single dust temperature along the line of sight, and a single dust-to-gas ratio (DGR) for the whole GMC,
(this method will hereafter be referred to as IR-derived; for more details about the method please see the observations and data section, Section ~\ref{sec:Obs_Data_3}).
By comparing H$_{2}$ and CO distributions, L12 estimated the fractional mass of ``CO-dark" H$_2$ (f$_{DG}$) within Perseus to be f$_{DG} \sim 0.3$.
This study also found that while H$_{2}$ is in general more extended than CO, significant spatial variations exist. 
As shown in Figure~\ref{f:Map}, $I_{\rm CO}$ and H$_2$ contours trace each other well on the west side, while
H$_2$ is significantly more extended on the east side.
However, as many assumptions are needed when deriving the IR-based H$_2$ distribution, and IR images provide integrated line-of-sight properties, comparing the estimated f$_{DG}$ with other methods is highly important.
This can be achieved by using the [CII] emission.

Another important reason for studying the 158 $\mu$m transition is its importance as the key cooling line for the cold ($<$ few $\times$ 100 K) interstellar medium at typical volume densities of a few$\times$10$^2$ cm$^{-3}$ (e.g. \citealt{1972ARA&A..10..375D}, \citealt{2003ApJ...587..278W}, \citealt{Tielens_Book}).
Under the assumption of thermal equilibrium, the intensity of the [CII] emission is indicative of the heating rate and provides information about the strength of the radiation field.



In this paper, we investigate properties of the [CII] emission in two PDRs in the Perseus molecular cloud as a case study for mapping out the transition from primarily atomic to primarily molecular regions.
Perseus is a molecular cloud that resides below the Galactic plane in the larger 
Taurus-Auriga-Perseus molecular complex. 
This GMC has a mass of $\sim$10$^4$ M$_{\odot}$, and is located at a distance of $\sim$300 pc with $\sim$30 pc difference between the east and west side, with Gould’s Belt Distances Survey distances for the western side near NGC 1333: $\sim293\pm22$ pc away and the eastern side near IC 348: $\sim321\pm10$ pc
\citep{1983Herbig, 2018ApJ...869...83Z, 2018ApJ...865...73O}.  
Its close proximity makes it a good candidate
for high resolution, multi-wavelength studies.
As discussed in \citet{2014ApJ...784...80L},
the Perseus GMC has reached chemical equilibrium, meaning that the timescale for formation of H$_2$ within the main body of Perseus is much shorter than the age of this GMC.
It is also important to note that Perseus has not formed many massive stars, with no O-type stars and only five B-type stars.

The key goals of this study are to: 
(1) map out the spatial extent of the [CII] emission in two boundary regions of Perseus and investigate the kinematics of the transition layer relative to the central regions; 
(2) compare integrated intensity profiles of [CII] and CO and investigate whether a steady-state, chemical equilibrium PDR model can reproduce these profiles; 
(3) and investigate which ISM phase the [CII] emission in Perseus is mainly associated with.
To accomplish these goals we obtained [CII] observations
with {\it Herschel} in two different regions within Perseus. 

One region, ``branch A'', is located near the active star-forming region NGC1333 and 
was found to have $f_{DG}\sim 0.2$ by L12 based on IR observations.
The other region, ``branch B'', probes a more diffuse portion of Perseus with 
$f_{DG}\sim 0$, as shown in Figure~\ref{f:Map}. 
By observing these two regions, one with and one without ``CO-dark" molecular gas, 
and comparing their [CII] emission we will also be able to compare estimates of the ``CO-dark'' H$_2$ gas from two different methods: the IR-based method and the [CII] method.



The structure of this paper is organized in the following way. In Section \ref{sec:Background_2} we summarize several previous studies of the ``CO-dark" molecular gas using the 158 $\mu$m [CII] emission.
Section \ref{sec:Obs_Data_3} details the observations and our data reduction procedures. The observational findings are displayed and discussed in Section \ref{sec:Results_4}. In Section \ref{sec:Model_5}, we summarize the PDR model by \citet{2010Wolfire} and use it to model [CII] and CO integrated intensity profiles. Finally in Section \ref{sec:Conclusions_6} we summarize the results of this paper.

\section{Background}
\label{sec:Background_2}

With its high angular and velocity resolution, the Heterodyne Instrument for the Far Infrared (HIFI) on board the {\it Herschel} has enabled detailed studies of the [CII] emission across different interstellar environments enabling investigations of the ``CO-dark" H$_2$ gas.
Several {\it Herschel} studies have suggested that the [CII] emission is a good tracer
of diffuse H$_2$ that does not coincide with bright CO emission (e.g. \citealt{2010Langer}, \citealt{2013Pineda}, \citealt{2015Gerin}). 
For example, \citet{2014Langer} used the GOT C+ survey (Galactic Observations of Terahertz C+) of the Milky Way plane to select 1804 individually detected [CII] components in the direction of  $\sim$150 sightlines.
This study showed that the [CII] emission that is not cospatial with any CO emission cannot arise entirely from the diffuse atomic medium, as the measured [CII] intensity is stronger than what can be produced by collisions with only hydrogen atoms, implying that the gas has a significant H$_2$ component.
\citet{2013Pineda} estimated that, for GOT C+ observations, 
less than 4\% of the total [CII] emission was associated with the
warm ionized medium.

While \citet{2014Langer} concluded that a significant amount of ``CO-dark" molecular gas can be traced using [CII] emission, they noticed that
the f$_{DG}$ varies across different phases and densities of the ISM. 
The diffuse molecular components have an average f$_{DG}$ = 0.4 and dense molecular components f$_{DG}$ = 0.2, making the average for their entire sample f$_{DG}$ $\sim$ 0.3.
L12 calculated a f$_{DG}$ $\sim$0.3 for the Perseus GMC,  by comparing the total H$_2$ mass enclosed within 3-$\sigma$ contours of the CO and H$_2$ distributions,
which is in close agreement with the statistical average \citet{2014Langer} calculated from single lines of sight.

While GOTC+ sampled [CII] emission across the Galactic plane, several
studies have investigated [CII] emission in individual GMCs. For example, \citet{2014Orr} analyzed a PDR across a boundary region in the Taurus GMC. 
As Taurus has a similar total mass \citep{2010ApJ...724..687L} and dust temperature \citep{2014A&A...571A..11P} to Perseus, it provides an important comparison point.
\citet{2014Orr}  found no significant [CII] emission for the region they observed, 
and their upper limits used with the Meudon PDR model \citep{2006LePetit} suggested a very low incident ISRF of 0.05\hspace{0.7pt}G$_0$, which was consistent with previous studies of Taurus \citep{2010Pineda}.
When changing different input parameters in the Meudon PDR code, \citet{2014Orr} found that the ISRF was crucial for explaining the [CII] intensity, while variations in suprathermal chemistry, inclination, and clumping did not have a significant contribution.

The 158 $\mu$m [CII] emission has also been observed in infrared dark clouds (IRDCs) and found to be almost anti-correlated with dense gas. This suggested that [CII] emission and C$^+$ are not spatially coincident with the densest parts of IRDCs and are more likely to be spatially correlated with more diffuse molecular or cool neutral gas \citep{2014A&A...571A..53B}.

Prior to the {\it Herschel} observations detailed in this paper, there have been several studies
of the [CII] emission in Perseus, however most early observations were affected by too-large a beam size or poor velocity resolution or both, making it difficult to resolve individual regions. 
For example, the Far Infrared Absolute Spectrophotometer instrument on-board the {\it Cosmic Background Explorer} satellite detected [CII] emission from Perseus, but had a beam size of $\sim7^{\circ}$ and a spectral resolution too low to resolve individual emission lines \citep{1994ApJ...434..587B}. 
This study concluded that the [CII] emission arose from the cold neutral medium (CNM). 
This conclusion was the result of calculations detailing the thermal pressure of HI necessary to produce the observed [CII] intensity. 
These calculations found that the [CII] emission could be explained by a medium with pressures between 1000 and 2000 cm$^{-3}$\hspace{0.9pt}K.

The Long Wavelength Spectrometer (LWS) instrument on board the {\it Infrared Space Observatory} satellite provided observations of Perseus at a higher angular resolution of $\sim1'$ but with a velocity resolution of $\sim1500$ km s$^{-1}$ \citep{2001A&A...379..557B}. 
\citet{2002Young} studied nine low-excitation reflection nebulae (including the reflection nebula produced by NGC 1333 within Perseus) using FIR observations from the {\it Kuiper Airborne Observatory} and incorporated these into PDR models.
They obtained observations of the [OI] 63 $\mu$m and 145 $\mu$m, [CII] 158 $\mu$m, and [SiII] 35 $\mu$m fine-structure lines. 
The line ratios provided estimates of the density, temperature, and incident ISRF.
For the NGC1333 nebula, using their observations in conjunction with the PDR model from \citet*{1991ApJ...377..192H}, they obtained a UV ISRF of 4800G$_0$ and a volume density of $2\times 10^4$ cm$^{-3}$.

\section{Observations and Data Processing}

\label{sec:Obs_Data_3}
\subsection{[CII]}
Observations of the fine-structure transition of C$^+$
($^2$P$_{3/2}$-$^2$P$_{1/2}$) at 1900.5369 GHz (rest frequency), were obtained 
with band 7b of the HIFI  instrument \citep{2010deGraauw} on-board {\it Herschel}
\citep{2010Pilbratt}. The [CII] spectra were obtained
using the Wide Band Spectrometer with 0.07887 km s$^{-1}$ velocity
resolution over 150 km s$^{-1}$.
For each target position two polarizations were recorded using the
Load CHOP (HPOINT) mode, with a sky reference at 1.4$^\circ$ to 2$^\circ$ off
from the target.


Figure 1 shows the {\it Herschel} observation positions, displayed on L12's 
HI integrated intensity image. 40 spectra were observed in  total, 
20 for ``branch A'' and 20 for ``branch B'' 
(Figure ~\ref{f:Map}).  Overlaid on this image are 3$\sigma$ contours for both the $^{12}$CO integrated intensity ($I_{\rm CO}$) from the COMPLETE survey \citet{2006Ridge} and \citet{2001Dame} and the H$_2$ surface density map derived from IR and visual extinction observations by L12. 
Both branches start inside Perseus, as traced by CO, and
extend toward its outskirts into lower-$A_V$. 
Branch A and branch B were selected to sample two PDRs with different observed ``CO-dark'' gas fractions.
Branch A is located where H$_2$ is found to be more extended than CO with a f$_{DG} \sim0.2-0.3$, while branch B
is in a region where H$_2$ and CO contours agree well with with a f$_{DG}\sim$0 \citep{2014ApJ...784...80L}.
Positions along both branches were distributed to match the resolution of the available $A_V$ 
observations of the region, with $\sim$ 4.3' between pointings\footnote{ At this declination, the physical distance between points is not strongly affected by projection effects.}.
This angular separation also roughly coincides with the resolution of the IR-derived H$_2$ column density map from L12. 

The spectra were reduced within the {\it Herschel} Interactive Processing Environment (HIPE) version 14.2.0
using the Heterodyne Instrument for the Far-Infrared (HIFI) pipeline \citep{2006ASPC..351..516O}. 
Reference spectra were not subtracted, due to a significant contamination from off source [CII] emission (see Appendix \ref{s:app-reduction} for tests on our reduction method). 
This may introduce significant error due to the prominence of several different types of standing waves, instrumental response, and drift. These standing waves include non-sinusoidal standing waves introduced by the Hot-Electron Bolometer (HEB) Mixer. 

During the reduction process, we first applied the ``hifipipeline'' command which 
brings raw data from level 0 to level 2.5. 
This task finds and masks bad pixels, accounts for the non-linear response of the CCDs, and derives the frequency range from the applicable comb spectra and then applies this frequency calibration, 
the transition to level 0.5 data.
A comb signal is
a series of stable frequencies at 100 MHz steps which is used to assign a frequency scale to the Wide Band Spectrometer (WBS) CCD channels.
During the transition to level 1 data, hot/cold load measurement standing waves are removed and each channel weight is calculated from these reduced form hot/cold load measurements. 
 
Level 1 data have also been temperature calibrated using channels' weights, velocity corrected to account for motion of the satellite, and a HEB standing-wave correction has been applied.
Band 7b of HIFI is well known to have prominent standing waves generated 
between the HEB mixer and the first low noise
amplifier. These waves are non-sinusoidal and mixed with typical, 
sinusoidal instrumental standing waves of periodicities $\sim$300 MHz.
The ``doHebCorrection'' task fits generated non-sinusoidal functions to the data, to correct for these electronic standing waves generated by the HEB for bands 6 and 7 of HIFI.
For more information on ``doHebCorrection'' task, please refer to Section 12.4 of the HIFI data reduction guide or one of the many 
papers published about HIFI by the {\it Herschel} team, e.g. \citet{2017A&A...608A..49S}.
After this, corrections for telescope dependent parameters are taken into account and output spectra are in units of antenna temperature. HIPE divides by sideband gain coefficients and stitches the three subbands together for each polarization.

\begin{figure*}
  \includegraphics[width=0.98\textwidth]{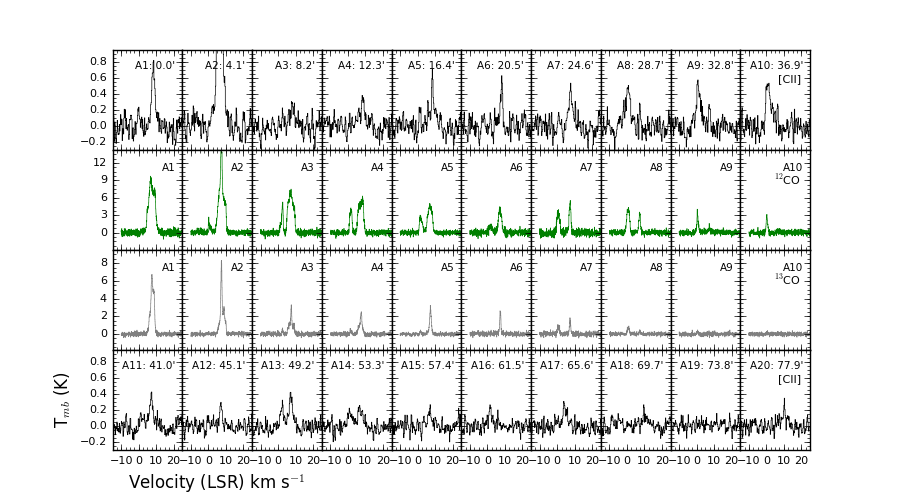}
  \caption{[CII], $^{12}$CO, and $^{13}$CO spectra for positions within the A branch in the range $-$15 km s$^{-1}$ to 25 km s$^{-1}$. 
  The [CII] spectra are in black with a velocity resolution of 0.08 km s$^{-1}$; 
  the $^{12}$CO(J=1-0) in green and $^{13}$CO(J=1-0) in gray, both with a velocity resolution of $\sim$0.07 km s$^{-1}$ \citep{2006Ridge}. 
  There is no significant $^{13}$CO emission for positions A11-A20 and no significant $^{12}$CO emission for positions A12-A20, 
  therefore the $^{12}$CO and $^{13}$CO spectra for A11-A20 have been omitted.}
  \label{f:AllASpectra}
\end{figure*}

However, the level 2.5 data had large standing waves due to the lack of reference spectra subtracted. 
To deal with this we used the ``fitHifiFringe'' task twice per polarization, fitting the same frequency standing waves for all pointings. 
The ``fitHifiFringe'' task combines sinusoidal functions of different periods to best fit the underlying standing wave structure contributed by all processes not related to the HEB.
This fitting procedure fit standing waves with a combination of sinusoids with periods of $\sim$95 MHz, $\sim$150 MHz, and $\sim$350 MHz for the vertical polarization and $\sim$45 MHz, $\sim$95 MHz, and $\sim$105 MHz for the horizontal polarization with a slight scatter around these frequencies for each individual spectrum. 
This scatter is introduced by the fitHifiFringe procedure which fits to minimize $\chi^2$ values.
The application of ``fitHifiFringe'' once is usual, but owing to the presence of many different instrumental effects, from the lack of off-spectra subtraction, two functions were necessary for the present data.
Any third attempt at applying ``fitHifiFringe'' resulted in fitted functions with amplitudes smaller than the standard deviation of the spectrum.
These were deemed extraneous and within random noise limits, halting the effectiveness of this procedure at 2 applications.
The function ``fitHifiFringe'' 
outputs a $\chi^2$ plot as a function of frequency of the fitted sinusoids.
The fitted sinusoids are located at the frequency of localized minima. 
Visual inspection of the $\chi^2$ ensures a higher probability that all prominent standing waves are being
subtracted. 
We restrain the fitting process to sinusoid frequencies $\>50$ MHz to avoid introducing narrow 
lines that may interfere with or be influenced by emission peaks.

After these functions were subtracted from both polarized bands, the vertically and horizontally polarized signals were combined into a final spectrum. The frequency scale was converted into a velocity scale, with the velocity origin corresponding to the rest frequency of [CII], 1900.5369 GHz.
Finally, following \citet{2014Orr} $\&$ \citet{2017Pineda}, 
a third-order polynomial was fit and subtracted from the HIPE output spectra 
using a simple Python, polynomial-fitting routine to flatten each spectrum's residual baseline structure. 
The polynomial-subtracted spectra were then smoothed over 5 channels
to a velocity channel width of 0.39 km s$^{-1}$.
To convert the data from antenna temperature ($T_{A}$)
to a main-beam temperature ($T_{mb}$) scale we divided by the empirically derived main beam efficiency: 
0.69 \citep{2012A&A...537A..17R}. The median standard deviation of smoothed spectra 
in the A branch is $\sigma_{A-smoothed}=0.06 K$ ($\sigma_{A-unsmoothed}=0.11 K$) and for the B branch is $\sigma_{B-smoothed}=0.05 K$ ($\sigma_{B-unsmoothed}=0.08 K$).

As a double check on our data reduction methods and calibration we compared the [CII] intensity 
(at {\it Herschel}'s resolution of $12"$) for 
the position corresponding to the NGC 1333 reflection nebula (A2) with previous observations
from \citet{2002Young} which had a resolution of $\sim30"$.
Their [CII] intensity is 4.8$\pm$0.1 (10$^{-4}$ ergs s$^{-1}$ cm$^{-2}$ sr$^{-1}$).
In the same units, the measured intensity is 5 $\pm$0.1 (10$^{-4}$ ergs s$^{-1}$ cm$^{-2}$ sr$^{-1}$), further solidifying the validity of our data processing methods.
An additional test of our methods is shown in Appendix~\ref{s:app-reduction}.

\begin{figure*}
	\includegraphics[width=0.98\textwidth]{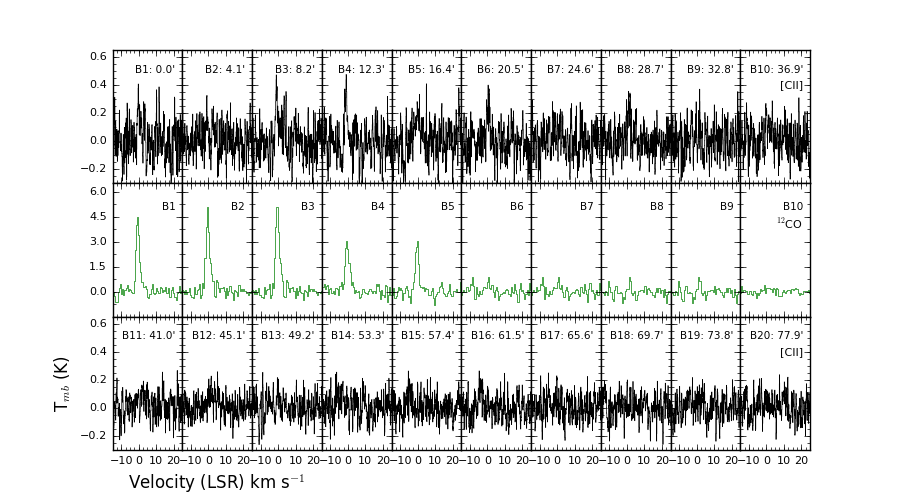}
	\caption{[CII] and $^{12}$CO spectra for the B branch. The [CII] spectra have a velocity resolution of 0.08 km s$^{-1}$ and the $^{12}$CO spectra have a velocity resolution of 0.65 km s$^{-1}$ \citep{2001Dame}. These spectra cover a range in local standard of rest velocity space of $-$15 km s$^{-1}$ to 25 km s$^{-1}$. The $^{12}$CO spectra for positions B11-B20 have no significant emission and have been omitted.}
	\label{f:AllBSpectra}
\end{figure*}

\subsection{Additional Data Sets (CO, HI, H$_2$, $A_V$)}
  \label{sec:Av}

We use the $^{12}$CO(J=1-0) and $^{13}$CO(J=1-0) spectra  from the COMPLETE Survey \citep{2006Ridge}. 
The datasets, respectively, have velocity resolution of 0.064 km s$^{-1}$ and 0.066 km s$^{-1}$ over a range of 40 km s$^{-1}$.
The $^{12}$CO and $^{13}$CO data sets have half-power beam widths of $46"$ and $44"$.
We used a main beam efficiency of 0.5 for 110 GHz and 0.45 at 115 GHz, respectively, to convert 
between $T_A$ and the main beam brightness temperature ($T_{mb}$) \citep{2006Ridge}. 
The $^{12}$CO and $^{13}$CO $T_{mb}$ data have a root mean squared (rms) noise of 0.35 K and 0.12 K per channel, respectively.
The COMPLETE survey does not have coverage past position A17 nor for any branch B positions.


For branch B the $^{12}$CO integrated intensities are obtained
from \citet{2001Dame}. \citet{2001Dame} produced a composite survey of 
the entire Galaxy at angular resolution of $8.4'$ by combining several 
different CO surveys of the  Galactic plane. The CO data were obtained with the 1.2-m telescope
at the Harvard-Smithsonian Center for Astrophysics. The
spectra were sampled with an angular spacing of $7.5'$ and the final
data cube for the Perseus region has a uniform rms noise of 0.25 K per a 0.65 km s$^{-1}$
channel. \citet{2001Dame} estimated the $I_{\rm CO}$, 
by integrating CO emission over the velocity range  $-15$ to 15 km s$^{-1}$ (see Section 2 of \citet{2001Dame} for further observation and analysis details).

The V-Band optical extinction ($A_V$) data used, is from the map released by the COMPLETE survey.
The COMPLETE team \citep{COMPLETE_2011} estimated the optical extinction from the 2-Micron All Sky Survey (2MASS) Point Source Catalog using the NICER algorithm \citep{2001A&A...377.1023L} with an angular resolution of $5'$ (for comparison with other data sets we investigate beam dilution effects of the $A_V$ image in Appendix~\ref{s:app-beam}). 
The NICER algorithm estimates the reddening along a line of sight by comparing the stars within the field to the intrinsic light of similar stars in a field with no reddening. 
This method involves no assumptions about dust or gas within a cloud and therefore gives the most unbiased estimate of the 
total dust column density \citep{2009ApJ...692...91G}.

The HI data are from Data Release 1 of the GALFA-HI survey with an angular resolution of $4.0'$, a velocity resolution of 0.18 km s$^{-1}$, and median rms noise of 0.19 K per velocity channel \citep{2011ApJS..194...20P}.
The HI column density image was created from the GALFA-HI data by integrating the HI brightness temperature over the velocity range of Perseus, $-$5 to 15 km\hspace{0.9pt}s$^{-1}$ and multiplying by  1.823$\times$10$^{18}$ cm$^{-2}$ (K km s$^{-1}$)$^{-1}$. 
\citet{2015Lee} corrected the HI column density for the high optical depth, we use their correction factors when dealing with the HI column density.

The H$_{2}$ map was derived in L12 from infrared IRAS and GALFA-HI observations.
First, IRAS images at 60 and 100 $\mu$m were used to derive dust temperature across Perseus, and the optical depth at 100 $\mu$m ($\tau_{100}$), assuming a single population of dust grains along the line of sight.  
Next, the optical depth image was converted to optical extinction A$_V$ 
by finding the conversion factor
which provides the best agreement between
our derived A$_V$ image and the V-Band optical extinction data from the COMPLETE Survey \citep{2006Ridge} for the region where the two images overlap.
Lastly, using an estimate of the local DGR of 1.1$\times$10$^{−21}$ mag cm$^{2}$, the H$_2$ column density was derived using the following equation: 
\begin{equation}
N(H_2) = \frac{1}{2}(\frac{A_V}{DGR}-N(HI)) .
\end{equation}

\subsection{Beam Dilution}

Due to differences in angular resolution between various datasets, beam dilution needs to be considered. 
We use the {\it Spitzer Space Telescope} 8 $\mu$m image (\citealt{2003PASP..115..965E, 2007Evans}) of Perseus to 
investigate beam dilution effects for the [CII] emission.
We also use the high resolution (18.2$\arcsec$) total column density image from the Gould Belt Survey \citep{2010A&A...518L.102A, 2012A&A...547A..54P} to consider beam dilution effects on the A$_V$ dataset in Appendix~\ref{s:app-beam}.
In general, we find that beam dilution is likely not a big effect for angular scales from $12''$ to $4.0'$ for both [CII] emission and A$_V$ datasets. 
This supports our comparison of integrated profiles in Section \ref{sec:Results_4}. 
We also emphasize that we are mainly focusing on large-scale trends of the [CII] integrated intensity, and do not attempt to investigate line ratios which would likely be more severely affected by beam dilution.

To probe the variation of CO intensity from pixel size of $46"$ or $44"$ to $4.1'$, we compared a Nyquist sampled distribution of positions spanning from A1-A20 to the sparsely sampled CO observations we use in the analysis. This comparison shows that the CO distribution between A1-A20 is reasonably well portrayed by the 20 pointings sampled at $4.1'$ (see Appendix~\ref{s:app-beam}).

\subsection{Gaussian fitting of individual spectra}

We fitted one or occasionally two Gaussian functions to each of [CII] and CO spectra (Figures~\ref{f:AllASpectra}
and ~\ref{f:AllBSpectra}).
The amplitude of the fitted Gaussian function was then compared 
to the noise level and each signal's detection was evaluated for statistical significance.
All [CII] detections in the A branch are over the significance level of 3$\sigma$. 
None of the B branch detections have a significance of $>$ 3$\sigma$, but there are 
several possible detections with significance $>$ 2$\sigma$.

We integrate the [CII] brightness temperature over the velocity range of $-5$ to 15 km s$^{-1}$.
Errors in integrated intensities were calculated by summing 
the channel errors in quadrature over the entire integrated velocity range. 
The errors in the central velocities and FWHMs of individual Gaussian components 
were generated from the covariance matrices created during Gaussian fitting.

\section{Results}
\label{sec:Results_4}

In this section we present the observations and show that the [CII] emission has a highly extended spatial distribution in Perseus (Section ~\ref{sec:ExEm}). We also investigate kinematics of Perseus as traced with [CII], $^{12}$CO, and $^{13}$CO line emission, to search for potential motions of the transition layers relative to the Perseus central regions (Section ~\ref{sec:Kine}). We then focus on comparing spatial trends of integrated intensities of [CII] and $^{12}$CO (Section ~\ref{sec:IntIntComp}), and investigate the ISM phase that the [CII] integrated emission is mainly associated with by comparing the [CII] intensity with the HI and total hydrogen column density (Section ~\ref{sec:HIH2CIIcomp}).

\subsection{Extended [CII] emission in Perseus}
  \label{sec:ExEm}

Figures~\ref{f:AllASpectra} and ~\ref{f:AllBSpectra} show the [CII] 158 $\mu$m, $^{12}$CO, and $^{13}$CO spectra (with $^{13}$CO for branch A only).
All CO spectra for positions beyond A10 and B10 have been omitted for space since only the $^{12}$CO A11 and A12 positions had significant emission.
The A branch has significant (SNR $>$ 3) [CII] emission detected in 19 out of the 20 positions, with a maximum brightness temperature of 0.8 K (excluding position A2), while the B branch has 12 positions that have a S/N $\geq$2.5 and a maximum 
brightness temperature reaching only $\sim$0.3 K.
All calculations of separation and physical scale are carried out under the assumption that Perseus is at a distance of 300 pc \citep{2018ApJ...869...83Z}.

The [CII] observations
qualitatively agree with the results from L12:
in branch A we detect significant [CII] emission in almost all positions, 
in agreement with the IR-derived H$_2$ suggestion that a significant amount of the ``CO-dark'' H$_2$ is present.
In branch B, which samples a more diffuse environment, much weaker [CII] emission is detected in
only 60\% of observed positions, in agreement with the expectation that 
no ``CO-dark'' H$_2$ gas is present there (based on L12).
This spatial comparison suggests, in agreement with several previous studies, that the [CII] emission could be a good tracer of the ``CO-dark'' H$_2$ gas and that ``CO-dark'' H$_2$ estimates using both the IR-derived H$_2$ and [CII] emission are in reasonable agreement.

The observed profiles also demonstrate that C$^+$ is present, at least, up to $\sim80'$ or $\sim$7.0 pc from the Perseus center, further confirming that the Perseus envelope is extended, and indicating that the observational sampling has not reached the edge of the diffuse envelope surrounding Perseus.
This is in agreement with L12's estimate, that Perseus has a highly extended envelope with both HI and H$_2$ being present up to $200'$ away from the centers of key star-forming regions.

Figure~\ref{f:AllASpectra} shows that
there is not only one spectral line, at $\sim8$ km s$^{-1}$, but for some positions within branch A there is
a second feature at $\sim$1 km s$^{-1}$. 
This emission peak is also seen in both the $^{12}$CO and $^{13}$CO spectra.
Considering that the $^{12}$CO velocity dispersion measured around NGC 1333 (e.g. \citealt{2014ApJ...784...80L})
is 1-2 km s$^{-1}$, we expect that this velocity component traces a dense clump that is associated with Perseus.
For direct comparison to $A_V$, as well as HI and H$_2$ maps, we assume that 
the two peaks trace molecular gas in Perseus and we
integrate the CO spectra over the same velocity range as for the HI, $-5$ to 15 km s$^{-1}$ 
(this range was initially estimated in L12).

\begin{figure}[htp]
\includegraphics[width=0.47\textwidth]{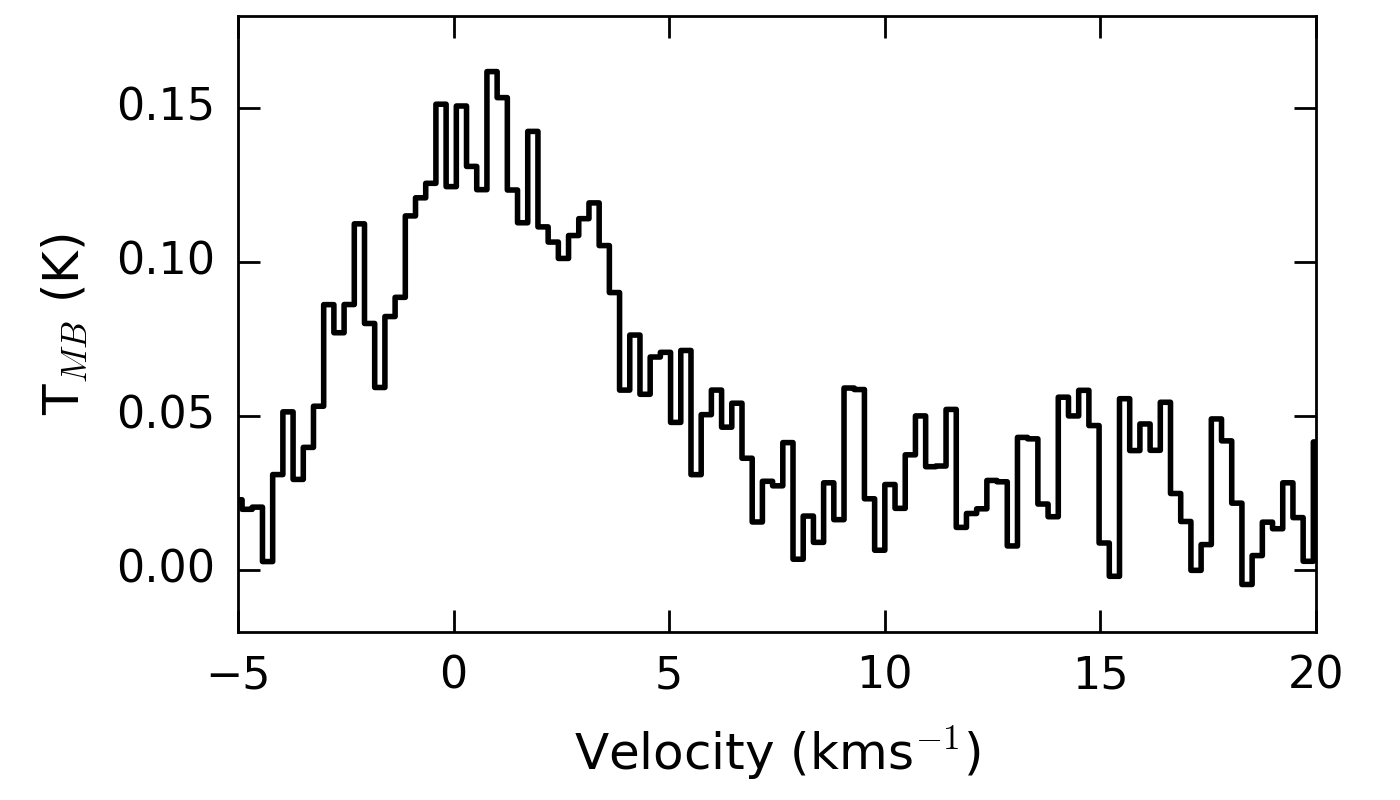}
\caption{A stacked spectrum obtained using [CII] spectra from B1-B20 positions of branch B. Each spectrum was smoothed over 5 channels to achieve a velocity resolution of 0.24 km s$^{-1}$.
Spectra were first shifted in velocity so that their peak is positioned at 0 km s$^{-1}$, 
then averaged together.}
\label{f:B-stacked}
\end{figure}

While there are fewer significant I$_{\rm [CII]}$ detections in branch B, and the peak emission is found with a central velocity of $\sim$1 km s$^{-1}$, it is possible that [CII] emission from other velocity components or positions 
is below the sensitivity. To test this hypothesis,
Figure~\ref{f:B-stacked} shows stacked spectra from positions B1-B20. 
To produce this spectrum we have shifted each spectrum in velocity so that its fitted Gaussian peak lines 
up at 0 km s$^{-1}$, and then added the spectra together and divided by the total number of spectra added, 20. 
As the figure displays, the stacked spectrum has a peak at 0 km s$^{-1}$ and a shoulder at 5 km s$^{-1}$. 
This suggests that although most points past B10 do not show significant emission individually, they likely each have [CII] emission but with an intensity below the noise level.

\begin{figure}[htp]
\includegraphics[width=0.45\textwidth]{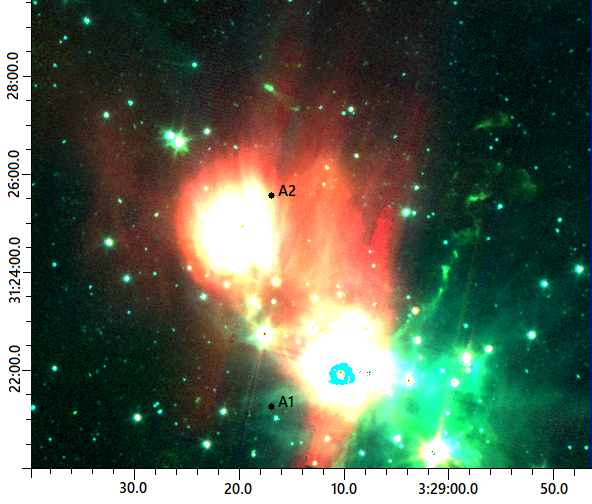}
\caption{{\it Spitzer} IRAC RGB image of the NGC 1333 region produced using GLIMPSE (Galactic Legacy Infrared Mid-Plane Survey Extraordinaire)  survey data with IRAC (3.6 $\mu$m) band 1 as blue, IRAC band 2 (4.5 $\mu$m) as green, and IRAC band 4 (7.9 $\mu$m) as red \citep{2004ApJS..154...10F, 2004ApJS..154..322C}.
This image displays a small field close to NGC1333 showing the reflection nebula associated with it.
The black circles correspond to positions at the base of branch A.
}
\label{f:Spitzer}
\end{figure}

We also note that the [CII] emission at position A2 is exceptionally bright with a peak $T_{mb}= 5.5$ K. 
A2 coincides with a dusty reflection nebula in NGC 1333 which has $A_V\sim8$ mag.
This dusty  nebula contains hundreds of young stars and is excited by UV photons and outflows. 
However, as evident in positions A1 and A3, the nebula is spatially small in diameter ($<$6$\arcmin$).
We verified this by examining the {\it Spitzer} images of the region: the reflection nebula encompasses less than 4 arminute$^2$, as is seen in Figure ~\ref{f:Spitzer}. 
While position A2 is within the brightest portion of the reflection nebula, our visual examination of the {\it Spitzer} images as well as our qualitative analyses of the {\it Spitzer } 8 $\mu$m and {\it Herschel} Gould Belt Survey high resolution column density image (see Appendix ~\ref{s:app-beam}) suggest that other positions are likely not affected by similar dusty structures.

The [CII] emission observed within the A branch is bright and essentially present in all pointings, extending
at least up to $82'$ away from the center of NGC 1333. 
At $\sim50'$ from NGC 1333 (pointing A13) there is 
no detected $^{12}$CO emission while there is still
significant [CII] emission.
As seen in Figure 2, the detected [CII] emission is spatially more extended than the detected $^{12}$CO emission, when taking into account respective noise levels.
Such extended and bright [CII] emission in Perseus contrasts with a similar study of the Taurus molecular cloud \citep{2014Orr} where at similar sensitivity [CII] emission was not detected. 

With a goal of understanding how interstellar environment affects [CII] emission, we note that bright [CII] emission was detected in regions of massive star-formation.
For example, using {\it Herschel} \citet{2015ApJ...812...75G} found a peak brightness 
temperature of $\sim$250 K  
in the Orion molecular cloud 1, which has a large incident ultraviolet radiation field arising from the close-by Trapezium cluster of young, bright stars
\citep{Bally2008}.
The Orion Bar was observed using {\it Herschel}, and yielded observations with a peak
T$_{mb}$ $\sim$75 K.
These previous observations suggest that the variability of the incident FUV ISRF, 
even on cloud scales, is of importance for the [CII] intensity.

\subsection{[CII] and CO kinematics}
  \label{sec:Kine}

We compare the central velocity of individual velocity components in order to investigate the kinematics traced by [CII] and CO and search for potential motions of the transition layer relative to the cloud center, as would be expected if the envelope was expanding away or contracting onto the cloud.


\begin{figure}
\includegraphics[width=0.47\textwidth]{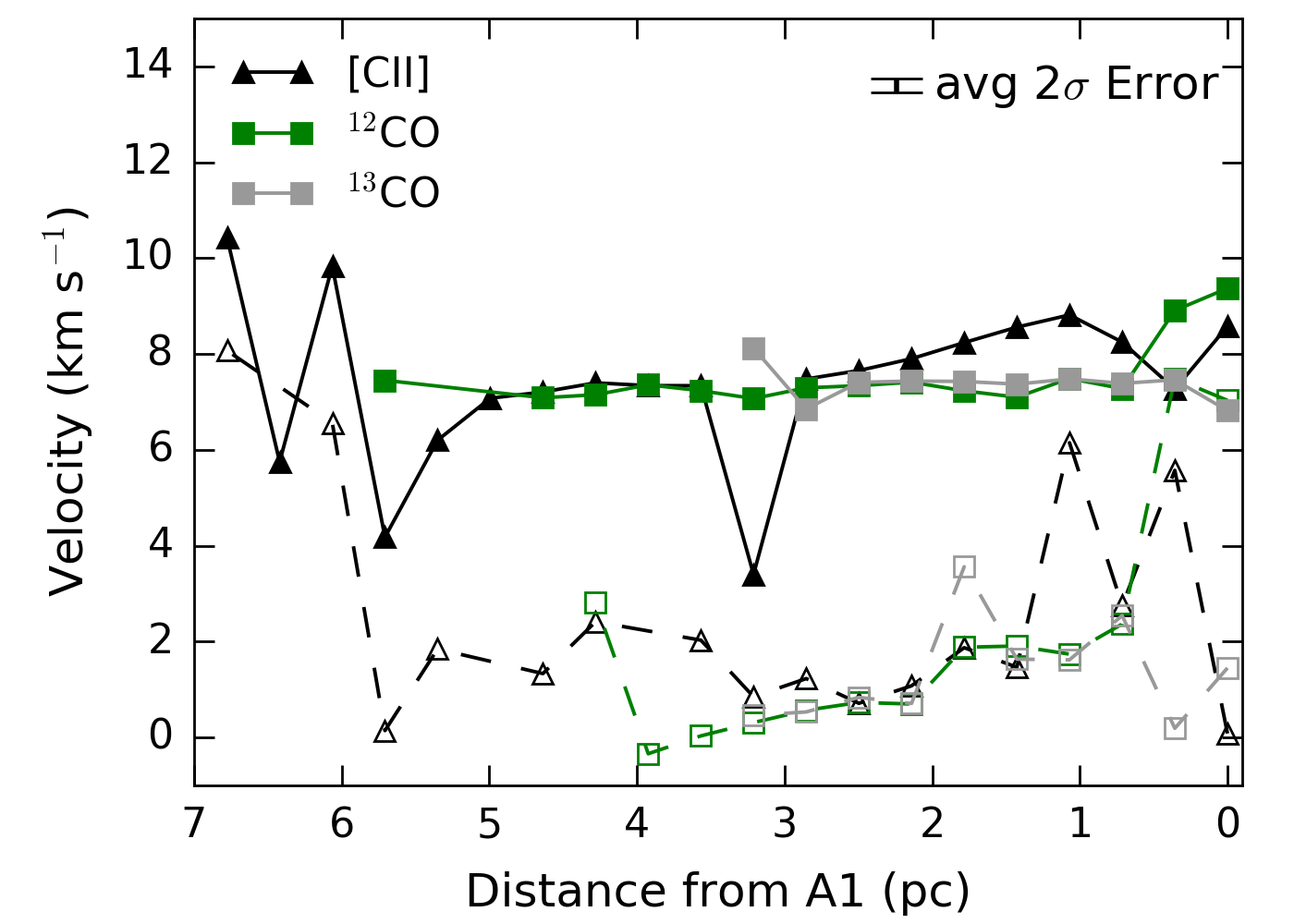}
\caption{Branch A central velocities of Gaussian components as a
function of distance from NGC1333 (in pc): [CII] as black triangles, $^{12}$CO as green squares, and $^{13}$CO as gray squares. Filled symbols correspond to the main component at $\sim$7 km s$^{-1}$ and unfilled symbols show the component at $\sim$1 km s$^{-1}$.  The typical 2$\sigma$ error of all tracers is indicated in the top right of the plot. We note that the errors for the first four points (A20-A17) of [CII] observations from the left are $\sim$2 times larger.
}
\label{f:7-cenvelocity}
\end{figure}

\begin{figure}
\includegraphics[width=0.47\textwidth]{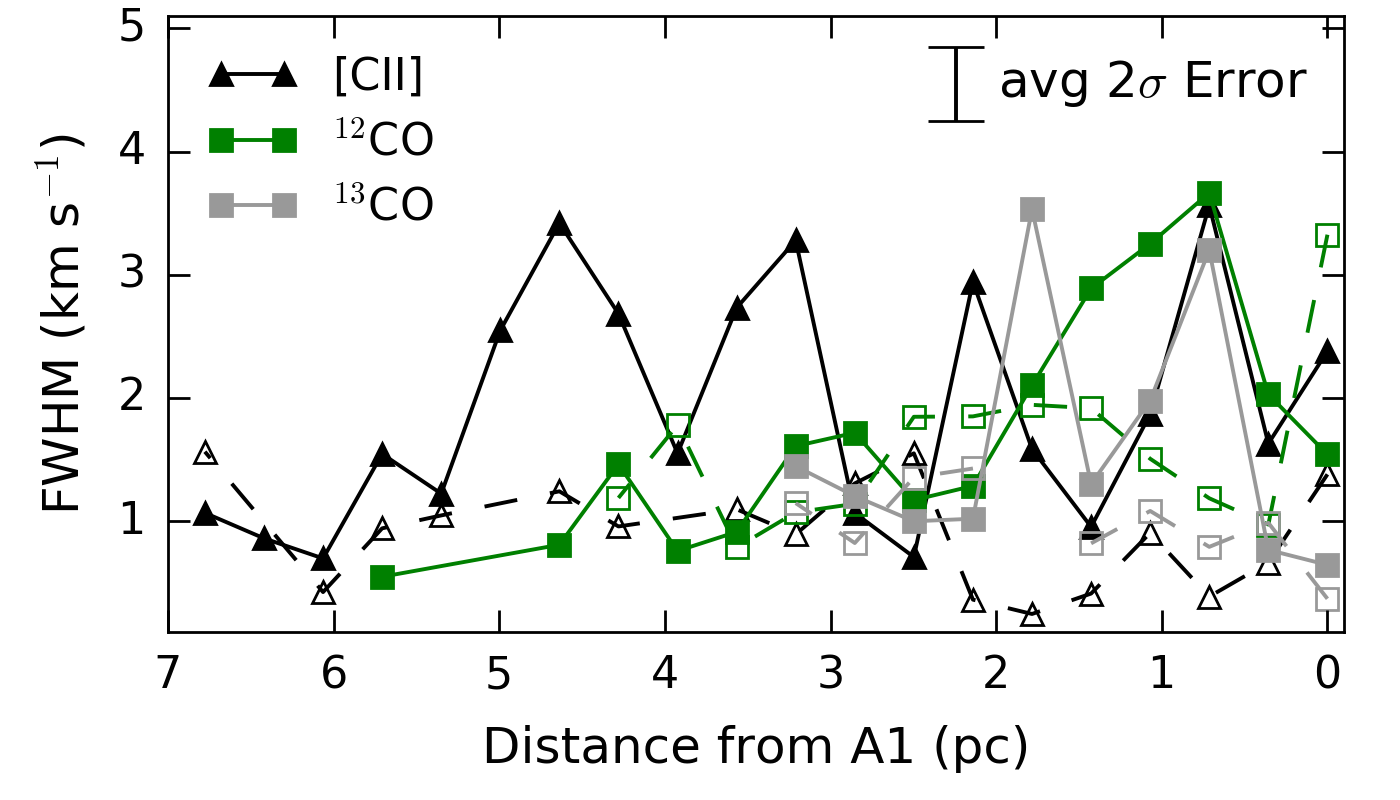}
\caption{Branch A velocity FWHM of Gaussian components as a
function of distance from NGC1333 (in pc): [CII] as black triangles, $^{12}$CO as green squares, and $^{13}$CO as gray squares. Filled symbols connected with solid lines correspond to the main component at $\sim$7 km s$^{-1}$ and open symbols connected by dashed lines show the component at $\sim$1 km s$^{-1}$.  The typical 2$\sigma$ error of all tracers is indicated in the top right of the plot. We note that the errors for the first four points of [CII] observations from the left (A20-A17) are $\sim$2 times larger.}
\label{f:7-FWHM}
\end{figure}

\begin{figure}[h]
\includegraphics[width=0.47\textwidth]{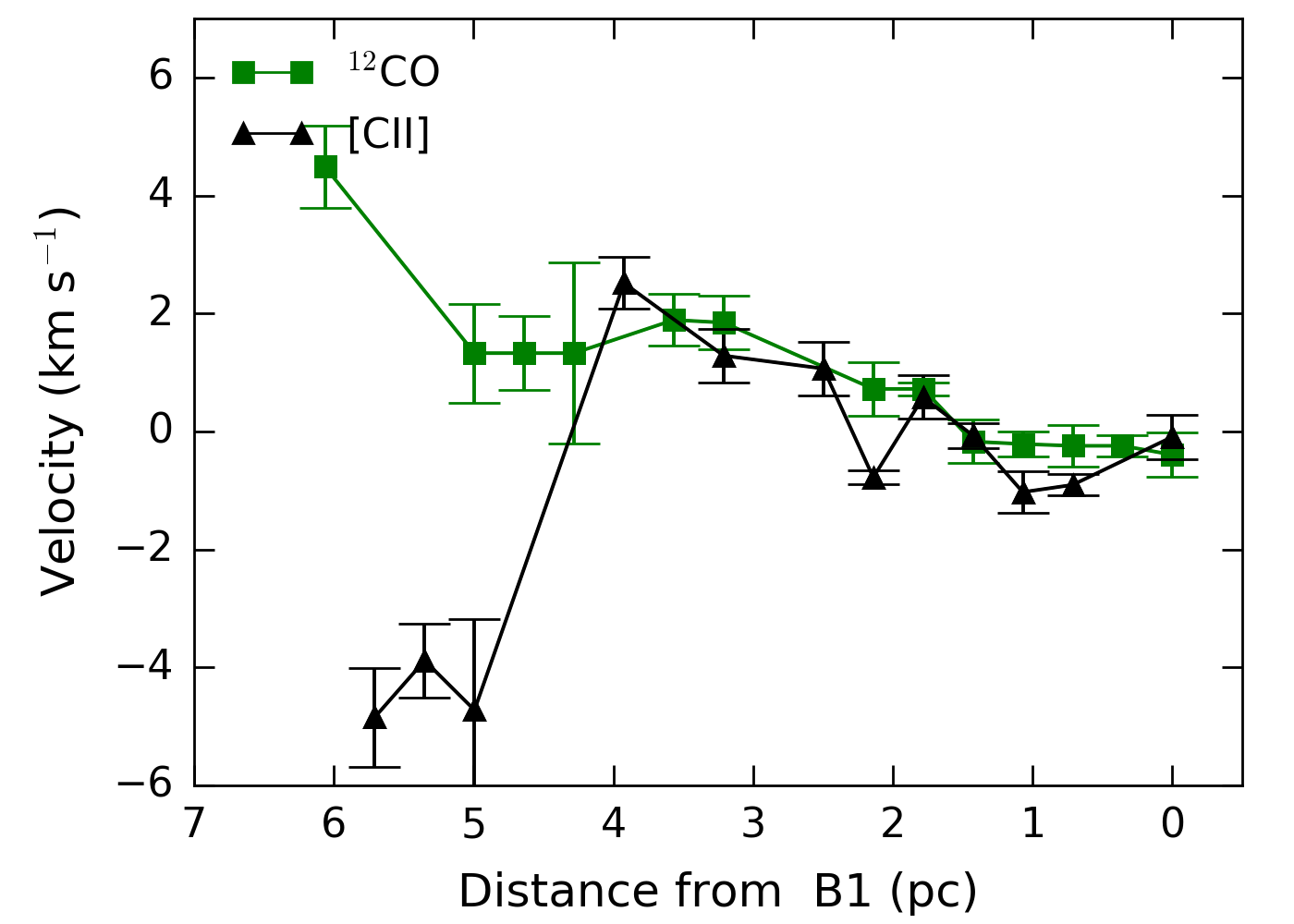}
\caption{Branch B central velocities of Gaussian components as a
function of distance from B1 (in pc): [CII] as black, $^{12}$CO as green.  }
\label{f:B_Cents}
\end{figure}

\begin{figure}[h]
\includegraphics[width=0.47\textwidth]{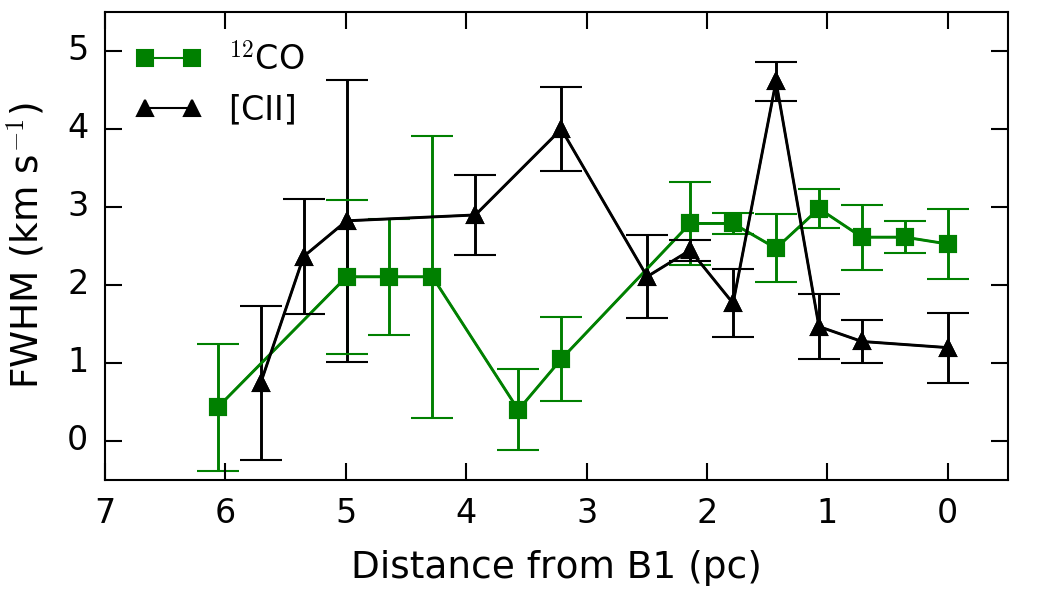}
\caption{Branch B velocity FWHM of Gaussian components as a
function of distance from B1 (in pc): [CII] as black, $^{12}$CO as green. 
}
\label{f:B_FWHM}
\end{figure}

Figures ~\ref{f:7-cenvelocity}, ~\ref{f:7-FWHM}, ~\ref{f:B_Cents}, and ~\ref{f:B_FWHM} show the central velocity and the velocity FWHM
of the [CII], $^{12}$CO, and $^{13}$CO ($^{13}$CO is available for branch A only) components from Gaussian fitting.
As seen in Figure~\ref{f:7-cenvelocity}, from $\sim$1 pc from NGC1333 to $\sim$5 pc away, both the central velocity of the $^{12}$CO and [CII] emission from the main body of the cloud remain consistent at $\sim$7 km s$^{-1}$ (within estimated uncertainties). 
There is a similar consistency of the central velocity of the component around $\sim$1 km s$^{-1}$, with a slight decline from $\sim$2 to 0 km s$^{-1}$ which is found in emission from the both CO isotopologues as well as C$^+$.
The last four [CII] pointings are less significant detections making Gaussian fitting more uncertain for both components, 
this results in a larger scatter of the central velocity. 
The lack of velocity shifts between [CII] and molecular tracers suggests that there are no clear motions of the transition layer relative to the cloud center. This is similar to what was seen in the case of L1599B \citep{2016Goldsmith}.

The central velocities of branch B (Figure ~\ref{f:B_Cents}), when compared to the central velocities of branch A, seem consistent with an overall decreasing velocity trend across the entire Perseus GMC.
\citet{1999ApJ...525..318P} saw a general trend of decreasing central velocity from east to west for the main body of Perseus, with an average central velocity of $\sim$12 km s$^{-1}$ near IC 348 on the eastern side, and $\sim$8 km s$^{-1}$ at NGC1333 (Branch A). 
Going $\sim$1.5$^{\circ}$ further west of NGC1333 the central velocity of $^{12}$CO emission is $\sim$4 km s$^{-1}$ and even further west, Branch B, which is 3$^{\circ}$ west of NGC1333, has an average central velocity of $\sim$1 km s$^{-1}$ (Figure ~\ref{f:B_Cents}) which fits into a continuation of the trend \citet{1999ApJ...525..318P} found. 
We note that while the central velocities of positions B15, B16, and B17 are closer to $-4$ km s$^{-1}$, they are still a part of Perseus with a significant HI component appearing at that position with a similar central velocity.
While velocity gradients across GMCs are not uncommon, their origin could be caused by any number of things such as rotation, shear, or expansion. As for branch A, except for the last few points which are uncertain, we find that [CII] and CO central velocities track each other well.

As PDR modeling and a consideration of optical depth (Sections \ref{sec:Model_5} and \ref{sec:IntIntComp}) require information about line widths, we also investigate
the FWHM of Gaussian fits.
On average the [CII] FWHMs from both branch A (Figure ~\ref{f:7-FWHM}) and branch B (Figure ~\ref{f:B_FWHM}) are
$\sim1$-2  km s$^{-1}$ with a large scatter from 1 to 3.5 km s$^{-1}$. 
The errors plotted are from the covariance matrix from the Gaussian fitting procedure.
The FHWMs of the CO components are systematically lower, typically around 1 km s$^{-1}$.
While at kinetic temperatures of 20-100 K the [CII] FWHM is expected to be slightly broader than the FWHM of CO (and below 1 km s$^{-1}$), both [CII] and CO linewidths are dominated by turbulent broadening.

\subsection{Comparison of [CII] and CO Integrated Intensities}
  \label{sec:IntIntComp}

\begin{figure*}
\includegraphics[width=\textwidth]{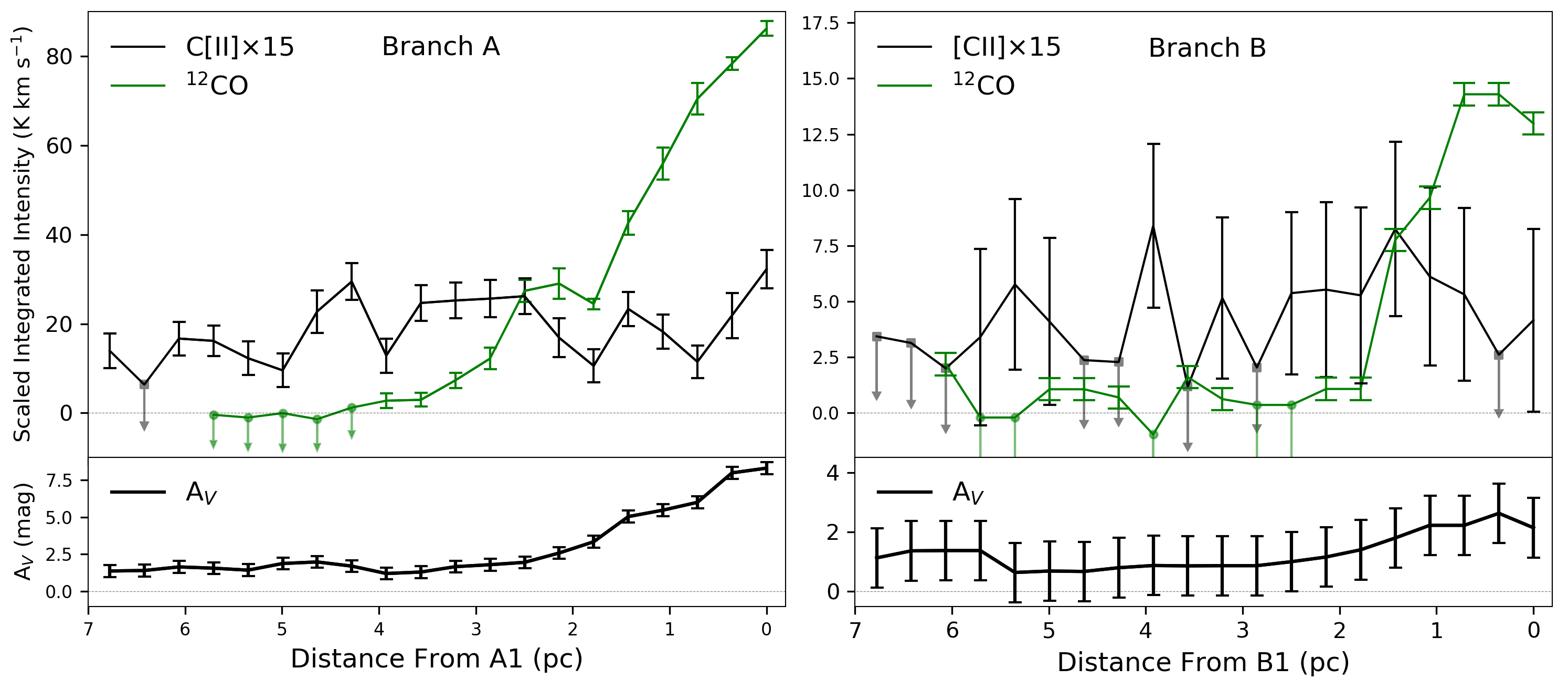}
\caption{Top Left and Right: The integrated intensity of [CII] and $^{12}$CO emission for as a function of distance from the first position of each branch (pc): [CII] data are plotted in black and $^{12}$CO(J=1-0) are plotted in green.  
Quantities for branch A are on the left and quantities for branch B are on the right.
All non-significant (SNR$<2.5$) positions are plotted as upper limits in arrows. 
Bottom Left and Right: $A_V$ plotted as a function of distance from the first position of each respective branch.
}
\label{f:IntCII_CO}
\end{figure*}

Figure~\ref{f:IntCII_CO} shows the integrated intensity of [CII] ($I_{\rm [CII]}$) and of $^{12}$CO ($I_{\rm CO}$), as well as $A_V$, as
a function of distance in parsecs from the first branch position within each branch, A1 in branch A corresponds with the center of star-forming region NGC 1333.
As the A2 position is affected by a reflection nebula, to estimate the intensity of the PDR region we replaced the A2 integrated intensities with the averages of points A1 and A3.\footnote{We note that the global results and modeling in the next section remain the same if we simply ignore A2 instead of replacing it with the average of A1 and A3.}
As seen in this figure, for branch A the $I_{\rm [CII]}$ essentially remains flat with a mean of 1.2 K  km s$^{-1}$ from NGC1333 until $80'$ (or 7.1 pc), while the I$_{\rm CO}$ rises with the rising accumulation of dust as shown by the $A_V$ profile. Similarly, $I_{\rm [CII]} \sim 0.3$ K  km s$^{-1}$ along the whole length of branch B, while $I_{\rm CO}$ rises close to the base of the branch.

\citet{2014ApJ...784...80L} and \citet{2008A&A...482..197P} investigated several star-forming and dark regions in Perseus and found that there is a threshold of dust corresponding to an A$_V \sim1$ mag necessary for shielding CO. 
We see a similar threshold in Figure~\ref{f:IntCII_CO}:
$I_{\rm CO}<3\sigma$ noise level for $A_V<1.2$ mag, for both branches A and B.
While branch B has a fewer number of significant detections (12 with SNR$>$2.5), its $A_V$ profile suggests a more diffuse environment relative to branch A, with a peak visual extinction of only 2.6 mag. 



The flat $I_{\rm [CII]}$ is different from what was found in a study of the boundary of a cloud envelope in L1599B by \citet{2016Goldsmith}. 
The [CII] observations of five points across the boundary showed that the [CII] intensity increased right at the cloud boundary.
To reproduce this observational trend through modeling, \citet{2016Goldsmith} needed a five times higher ISRF on one cloud side, in the direction of an O8 star. 
As a consequence of the enhanced ISRF, one side of the cloud envelope was warmer and the 
C$^+$ layer was much thicker and subsequently the [CII] emission was brighter.

Based on the study of L1599B, we would expect that $I_{\rm [CII]}$ would 
eventually drop for observed positions further from the Perseus center, 
yet we do not observe this.
For densities less than the critical density of the 158 $\mu$m [CII] transition (3800 cm$^{-3}$ and 3000 cm$^{-3}$ for collisions with H atoms, and 7600 cm$^{-3}$ and 6100 cm$^{-3}$ for collisions with H$_2$ at 20 K and 100 K respectively \citealt{2012ApJS..203...13G}),
[CII] line emission is the dominant gas coolant. 
Within this scenario,
all of the energy that goes into gas heating comes out in 
the [CII] 158 $\mu$m line emission. For a constant N(C$^+$) the gas heating integrated along the line of sight
depends on the incident FUV radiation field intensity 
and the photoelectric heating efficiency (the ratio of energy which goes into gas heating divided by the FUV photon energy; \citealt{1985ApJ...291..722T}). 
Under the assumption of thermal equilibrium, the observed flat $I_{\rm [CII]}$ 
profiles for both branches therefore suggest a constant heating rate all the way to $\sim80'$ ($\sim$7.0 pc) from the Perseus center. 

As the heating efficiency is relatively constant at $<\sim 3$-4\% \citep{1985ApJ...291..722T},
this implies a uniform incident radiation field. This result agrees with 
the uniform ambient radiation field around Perseus estimated by L12 using dust temperature as a proxy. In addition, the observed difference between branches A and B suggests roughly a factor of two higher heating rate in branch A relative to branch B. 
A factor of two higher radiation field, that can explain the 
higher heating rate in branch A,
would result in only a slight, almost indistinguishable, change in dust temperature for branch B and would still be consistent with the findings of L12.




The profiles, in particular for branch A, probe a significant range in terms of $A_V$: from $\sim1$ mag to $\sim10$ mag, yet the [CII] intensity remains uniform. Based on equation (3) from \citet{2018ApJ...856...96G}, the [CII] intensity, for densities well below the critical density, is proportional to temperature and volume density:
\begin{equation}
    I_{\rm [CII]} \propto  N(C^+) \times \exp{(-91.21/T_K)} \times T_K^{0.14} \times n(H)
\end{equation}
Assuming a constant N(C$^+$), this equation shows the relation between $T_K$ and $n(H)$ within the Perseus envelope. Under the assumption of uniform heating, in more central regions where
the density $n(H)$ (or $n(HI+2H_2)$) increases, based on the above equation, $T_K$ decreases.
In the outer regions, the opposite happens, as the density decreases, $T_K$ increases, resulting again in a relatively constant  $I_{\rm [CII]}$.

By using observational contraints for kinetic temperature, we can estimate the density needed to explain the intensity of [CII] emission, e.g. \citet{2018ApJ...856...96G}.
If we assume that the C$^+$ distribution is uniform throughout the Perseus envelope\footnote{This assumption is based on \citet{2018ApJ...856...96G}'s results
where N(C$^+$) was found to be fairly constant for a range of [CII] intensities:  N(C$^+$) = 1.1.-1.9$\times$10$^{17}$cm$^{-2}$ for a range of [CII] intensities: 0.160-0.681 K km s$^{-1}$.}, and that kinetic temperature and density are $n_1$, $T_1$ and $n_2$, $T_2$ in the outer and inner regions of the envelope, equation (1) and the observed uniform $I_{\rm [CII]}$ result in:
\begin{equation}
\frac{n_2}{n_1}=\left (\frac{T_1}{T_2} \right )^{0.14} \exp{\left (\frac{91.21(T_1-T_2)}{T_1 \times T_2} \right ) } 
\end{equation}
Assuming typical CNM conditions in the outer regions of the envelope with
$T_1=100$ K and $n_1=40$ cm$^{-3}$, and a temperature gradient such that $T_2=20$ K, we estimate the density in the central regions to be $n_2\sim2000$ cm$^{-3}$.
These rough estimates are in agreement with our results from PDR modeling in Section \ref{sec:Model_5}.

Finally, we note that a significant optical depth of the [CII] transition could result in a flat distribution of $I_{\rm [CII]}$
(e.g., \citealt{2013A&A...550A..57O}). To test for this possibility we use equation (2.68) from \citet{Tielens_Book}
to estimate the column density required to reach a line-averaged optical depth of unity. By assuming a line FWHM of 2 km\hspace{0.9pt}s$^{-1}$ (based on Figure 5) and a carbon abundance of $10^{-4}$cm$^{-3}$, we estimate
that a total, N(HI) + 2N(H$_2$), hydrogen column density of $1.4 \times 10^{22}$ cm$^{-2}$ is needed. 
This is significantly higher than the total hydrogen column densities we probe (discussed in the next section) in branches A and B, and we conclude that this scenario is not very likely.

\subsection{[CII] and HI, H$_2$ Comparison}
  \label{sec:HIH2CIIcomp}

\begin{figure*}
\includegraphics[width=\textwidth]{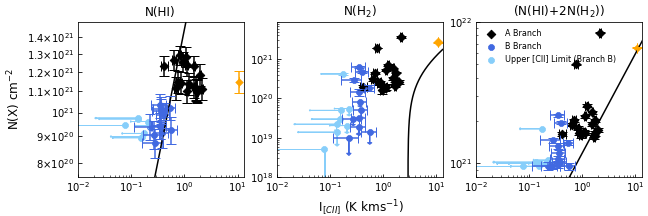}
\caption{The gas column densities plotted as a function of $I_{\rm [CII]}$ with branch B data in royal and light blue and branch A data in black. All significant values are plotted with double-sided, 2$\sigma$ error bars. Several non-significant values of $N({\rm H_2})$ are plotted as arrows indicating an upper limit for those positions. 
Left: The HI column density ($N({\rm HI})$) calculated from the GALFA-HI survey for the velocity range of Perseus ($-5$ to 15 km s$^{-1}$) and corrected for optical depth using the method from \citet{2015Lee} \citep{2011ApJS..194...20P}. Center: The H$_1$ column density, $N({\rm H_2})$, derived from IR observations by L12. Right: The total H column density ($N({\rm HI})$+ 2$N({\rm H_2})$). All light blue arrows are indicating upper limits of the integrated [CII] intensity. The orange diamond indicates position A2, which is within a reflection nebula. The black lines for the 1st and 3rd panels are fits of simulated data from \citet{2018Franeck}. The black line within the 2nd panel is the trend for the total H column density from \citet{2018Franeck} minus their trend for only the $N({\rm HI})$ and divided by 2.
}
\label{f:HI_CII_Compare}
\end{figure*}


While several previous numerical simulations suggested that 60-80\% of the [CII] intensity originates from regions dominated by molecular gas (e.g. \citealt{2017MNRAS.464.3315A, 2017Bisbas}), \citet{2018Franeck} proposed recently that for a newly formed GMC before the onset of massive star formation, up to 80\% of the [CII] emission could originate primarily from the CNM. 
\citeauthor{2018Franeck} used simulations of individual GMCs produced by the SILCC-Zoom project and applied 
a non-LTE radiative transfer model, RADMC-3D, 
to produce [CII] emission maps for individual GMCs without considering radiative feedback processes.
They found that the total gas column density as well as the HI column density, correlate with $I_{\rm [CII]}$. 

However, they concluded that [CII] is not a suitable tracer of the ``CO-dark'' H$_2$ gas for young GMCs at an evolutionary time of 13.9 Myrs because the dominant form of hydrogen is still atomic. 
For more chemically evolved GMCs it is likely that a larger fraction of the [CII] emission is produced in the H$_2$-dominated gas.
Considering that Perseus is $\sim10$ Myrs old, based on stellar ages, see L12 for discussion, and has no O-type stars and 3 B-type stars (most massive B5), it is reasonably similar to GMCs simulated by \citeauthor{2018Franeck}
Therefore within this section, we use the simulation predictions to investigate possible correlations between $I_{\rm [CII]}$, $N({\rm HI})$ and $N({\rm H_2})$ with a goal of investigating the origin of the [CII] emission.

Figure ~\ref{f:HI_CII_Compare} displays the HI column density, the IR-derived H$_2$ column density (from L12), and the total H nucleus column density ($N({\rm HI})$ + 2$N({\rm H_2})$) each as a function of $I_{\rm [CII]}$.
The black lines within the left and right panels of Figure ~\ref{f:HI_CII_Compare} are equations (9) and (8) from \citet{2018Franeck}, which are the fits of their simulated data. 
The black line within the middle panel is equation (8) minus equation (9), or the total gas column density 
minus the HI column density, leaving a proxy for just the H$_2$ column density.

We have corrected the HI column density for high optical depth using equation (15) from \citet{2015Lee}.
This correction is relatively small, and ranges from 1.09 to 1.14 with a median of 1.11 for the branch positions within our study.
Across the entire area of Perseus the correction factor reaches a maximum value of 1.2 as discussed within \citet{2015Lee}.
All A branch positions have significant $N({\rm HI})$ and $N({\rm H_2})$ values, while the B branch has 14 points with non-significant values of $N({\rm H_2})$.
As there is a significant difference in beam sizes, between {\it Herschel}'s $12''$ at 158 $\mu$m and Arecibo's $\sim4.0'$ at 21 cm, we considered the effect of beam dilution in the Appendix~\ref{s:app-beam} 
and concluded that this is not significantly affecting our comparison.

Figure ~\ref{f:HI_CII_Compare} shows that the two branches in Perseus probe regions with very different environments. 
By looking at all data points we see that observed $N({\rm HI})$ and $I_{\rm [CII]}$ 
agree reasonably well with the simulation prediction.
The two branches are seen as two distinct groups within this panel and have different median $I_{\rm [CII]}$ and $N({\rm HI})$ values of 1.0 K km s$^{-1}$ and 10$^{21.05}$ cm$^{-2}$ for branch A and 0.4 K\hspace{0.9pt}km\hspace{0.9pt}s$^{-1}$ and 10$^{20.9}$cm$^{-2}$ for branch B.
For the range of $I_{\rm [CII]}$ we probe, \citet{2018Franeck} predict the HI column density to be in the range of $10^{20.9-21.2}$ cm$^{-2}$, which is close 
to what we measure, but all of the positions, even those probing regions with high A$_V$ values, have $N({\rm HI})$ within the lower half of this range.
For the total gas or total hydrogen column density we observe a broader range of $10^{21.0-21.3}$ cm$^{-2}$ for most of the positions, with two positions reaching $\sim10^{22}$ cm$^{-2}$, generally higher than what the \citet{2018Franeck} simulations predict for the observed range of [CII] integrated intensity. While there is a larger scatter relative to the left panel, the total gas column density and the [CII] integrated intensity are not far off from the simulation predictions.

\begin{figure}
\includegraphics[width=0.47\textwidth]{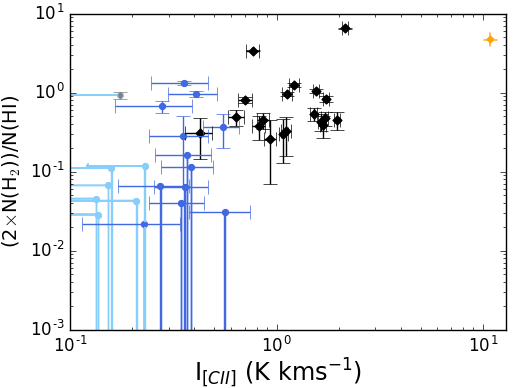}
\caption{2$\times N(H_2)$/$N({\rm HI})$ plotted as a function of $I_{\rm [CII]}$. Following the same notation scheme as figure ~\ref{f:HI_CII_Compare}, branch A positions are plotted as black diamonds, branch B positions with significantly detected $I_{\rm [CII]}$ are plotted as blue circles, and branch B positions with non-significantly detected $I_{\rm [CII]}$ are plotted as light blue circles with arrows indicating an upper limit. Branch B positions with non-significant $N({\rm H_2})$ are plotted as upper limits indicated by arrows instead of double-sided error bars.} \label{f:H_ratio}
\end{figure}

However, while the IR-derived $N({\rm H_2})$ is in the range of what is predicted by simulations, it corresponds to $\sim10$ times lower $I_{\rm [CII]}$ than is predicted by simulations. This results in a clear offset of observed points relative to the predicted relation in the middle panel of Figure ~\ref{f:HI_CII_Compare}. 
Again, there is a clear distinction between the A branch positions and the B branch positions, indicating that 
the B branch observations are probing a more diffuse region with about a factor of two lower $N({\rm HI})$, and in most cases more than a factor of two less $N({\rm H_2})$ as well as total gas density relative to the branch A. 
The exceptions are six positions within branch B with significant $N({\rm H_2})$ and a mean $2N(H_2)$/$N({\rm HI})$= 0.37.
The other 14 positions within the B branch have a factor of 10 lower $N({\rm H_2})$ than the mean significant values of $\sim$10$^{20.3}$cm$^{-2}$, while the A branch has two positions with a factor of 10 higher $N({\rm H_2})$.
$N({\rm H_2})$ has a much larger range suggesting that the H$_2$/HI ratio also varies significantly across individual branches and is overall higher in branch A (see Figure~\ref{f:H_ratio}).



Overall, the HI and the total Hydrogen column densities are in a good agreement with the predictions from \citeauthor{2018Franeck}, suggesting that
a significant fraction of the [CII] emission is likely associated with neutral gas from the Perseus HI envelope.
This agrees with the relatively uniform and extended [CII] distribution that reaches all the way to $\sim80'$ ($\sim$7.0 pc) from the Perseus center. 
While Perseus has reached chemical equilibrium \citep{2014ApJ...784...80L}, it does contain a significant amount of HI, with HI dominating the total mass budget. Also, Perseus lacks massive star formation and therefore its properties (HI and total gas column density) appear more similar to young GMCs simulated by \citeauthor{2018Franeck}
This comparison also suggests that Perseus is still in the process of converting a significant amount of HI gas into H$_2$ gas. Whether this HI is from the original GMC reservoir, or it was recently accreted, is unknown.


In summary, Figure ~\ref{f:HI_CII_Compare} suggests that a significant fraction of the [CII] emission is associated with HI and that HI clearly plays an important role in explaining the [CII] integrated intensity in Perseus.
The IR-derived H$_2$ column density is generally higher than the \citeauthor{2018Franeck} simulation predictions. Either the simulation is missing some H$_2$, or the IR-derived H$_2$ is possibly overestimated. 




\section{PDR Model}
\label{sec:Model_5}

PDR models are powerful diagnostic tools to examine the physical and chemical conditions within molecular clouds under the influence of a FUV radiation field.
We use here the PDR model by \citet{2010Wolfire} to investigate whether the observed [CII] and CO spatial trends can be explained under the assumptions of a steady-state chemical and thermal equilibrium. 
By examining the output parameters, integrated intensity of the $^{12}$CO and [CII] emission for each position (based on the input $A_V$ and the line FWHM), as well as gas temperature, we can gauge the strength of the local radiation field for two regions in Perseus.


\begin{figure*}
\includegraphics[width=\textwidth]{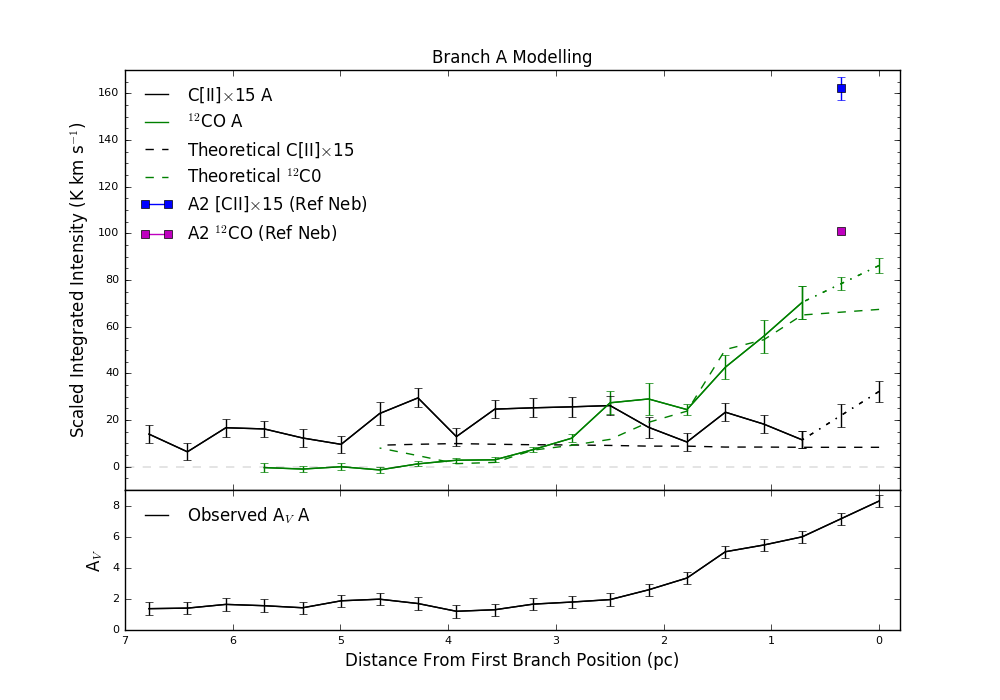}
\caption{Top: Plot of integrated intensities ($I_{X}$) as a function of distance from the first position of branch A. $I_{\rm [CII]}$ data are plotted in black, $I_{\rm CO}$ in green, observed quantities are plotted with solid lines and theoretically predicted values are plotted with dashed lines. Due to position A2 corresponding to a reflection nebula, the plotted $I_{\rm [CII]}$ and $I_{\rm CO}$ values for position A2 used within the trend are an average of the integrated intensities of A1 and A3. This point is connected to the others by dotted lines to indicate this assumed trend. The observed $I_{\rm [CII]}$ for A2 is plotted as a blue square and the observed $I_{\rm CO}$ as a magenta square. Bottom: $A_V$ plotted as a function of distance from the first position of branch A.}
\label{f:A_Theory}
\end{figure*}

\begin{figure*}
\includegraphics[width=\textwidth]{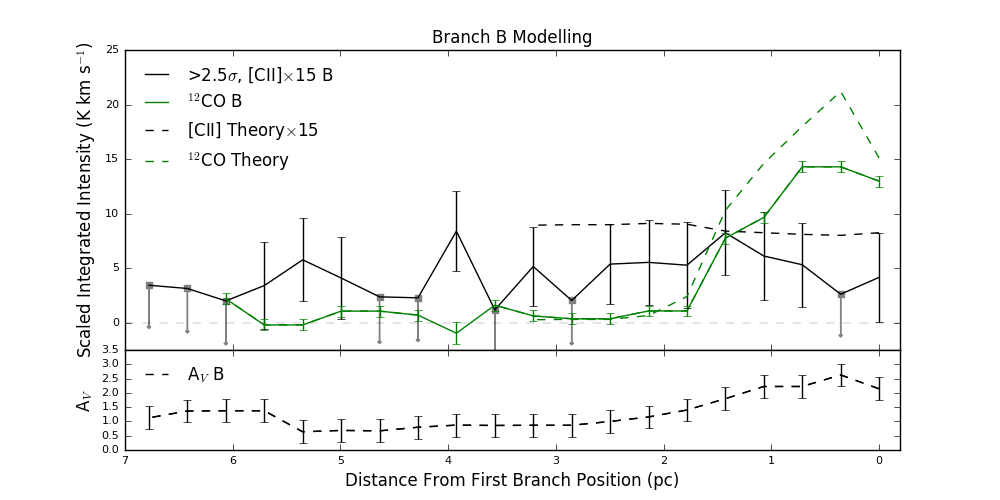}
\caption{Top: Plot of integrated intensities as a function of distance from the first position of branch B. $I_{\rm [CII]}$ data plotted in black with $>$2.5$\sigma$ detections as points with error bars and all $<$3$\sigma$ detections as upper limits plotted as arrows. The $^{12}$CO(J=1-0) integrated intensity ($I_{\rm CO}$) is in green. Observed quantities are solid lines and theoretically predicted values are plotted with dashed lines. Bottom: Dust extinction portrayed in $A_V$ is plotted as a function of distance from position B1 (pc).}
\label{f:B_Theory}
\end{figure*}

\subsection{Model Description}
We use a one-dimensional, plane parallel PDR model described in
\citet{2010Wolfire}, \citet{2012ApJ...754..105H}, and \citet{Neufeld2016}.
Here we update the photodissociation and photoionization rates according 
to \citet{Heays2017}. For the dependence of the rates with depth into 
the cloud we use the tabulated values for the 2nd-order exponential
integral function as appropriate for an isotropically incident radiation field.
The model assumes two-sided illumination of the plane and calculates the steady-state chemical
abundances and the gas temperature in thermal equilibrium as
a function of depth in the layer.

The density distribution assumed in the model is the same as that used in \citet{2014ApJ...784...80L}, namely 
an extended low density HI region (which we call the HI halo) surrounding a higher density region (which we call the core) of HI and H$_2$. This distribution successfully and simultaneously fitted N(HI), N(H$_2$),
A$_V$, and the $^{12}$CO (J=1-0) line intensity, as shown in \citet{2014ApJ...784...80L}. 
A low density was required to
match the N(HI) without converting the atomic gas to molecular,
while a high density was required to match the CO line intensities.
The exact density distribution is not well constrained beyond these specifications and an ad-hoc core halo model was adopted, the predictions made using this density distribution closely matched the observations. 
In the current paper we chose to use the same density
distribution to test whether the previous model could predict the [CII] observations as well as the $^{12}$CO integrated intensities presented here.

The input parameters for the model are: $Z$, DGR, $\xi$, $\chi$, $n$, $v_{D}$, and $A_V$, where $Z$ is the 
gas-phase abundance of elements, DGR is the dust to gas ratio, $\xi$ is the primary  cosmic-ray ionization rate per hydrogen atom, $\chi$ is the incident radiation field 
strength in units of the \citet{Draine1978} field\footnote{The Draine field is 1.69 times stronger than the Habing field and equal to $8.94 \times 10^{-14}$ erg cm$^{-3}$ 
for the integrated range of UV-radiation at 6-13.6eV \hspace{0.0pt} \citep{2011piim.book.....D}.}
, $n$ is the density of hydrogen nuclei, $v_D$ is the microturbulent Doppler line width ($=$FWHM/1.665), and
$A_V$ is the visual extinction through the layer.
To estimate these values we consider the physical properties of the Perseus GMC from the literature.
We use metallicity and DGR values as used in \citet{2014ApJ...784...80L}: 
$Z = 1 Z({R_\circ})$,  where 
$Z(R_\circ)$ is the gas phase abundance of elements at the
solar circle, and DGR = 1$\times$10$^{-21}$ mag\hspace{1.5pt}cm$^{2}$. 
For the cosmic-ray ionization rate we use $\xi=2\times 10^{-16}$
${\rm s}^{-1}$ from \citet{Neufeld2017}. For $\chi$ we assume the cloud is illuminated by
the interstellar radiation field and use a value of 0.5 incident on each side of the layer for the A branch
but find we need a factor of 2 lower to match the [CII] line emission
seen in the B branch. 
L12 estimated the incident ISRF for Perseus to be  equal to 0.4 Draine fields.
We use the observed FWHM measured for each position (Section \ref{sec:Kine}) which gives $v_D$, and takes into account both thermal and turbulent line broadening.
Finally, we use $A_V$ from the image released by the COMPLETE survey as described in Section ~\ref{sec:Av}.

In the \citet{2014ApJ...784...80L} representation, the core of H$_2$ and HI has a density of $n=10^{3}$ cm$^{-3}$ with a typical thickness of $\sim$1 pc 
(this density is motivated by several observational studies, e.g. \citealt{2008A&A...482..197P})
and is surrounded by an extended HI halo with a density of $n = 40$ cm$^{-3}$ and a depth of $\sim$3.5 pc. 
This density distribution reproduces the observed average HI column density of $N({\rm HI})$=9$\times 10^{21}$ cm$^{-2}$ with half contributing on each side of the slab. In the model,
there is no density gradient and the density changes abruptly between the halo and core once an approximate depth of $\sim$3.5 pc is reached.
The core size varies as a function of $A_V$ (across different positions), and under the assumed constant DGR, a larger $A_V$ value effectively creates a larger core.  
The core density needs to be $\geq$10$^3$ cm$^{-3}$ (i.e. the critical density 
of $J=1-0$ CO) to produce the $^{12}$CO and $^{13}$CO emission observed \citep{2014ApJ...784...80L}.
The model is not sensitive to the density of the extended halo as long as it is sufficiently diffuse enough to contain little H$_2$ and $^{12}$CO.

We note that with our density
distribution consisting of a low density halo and
separate high density core, we are essentially using
a simple clumpy model for the cloud in which
we are neglecting the emission and opacity from
interclump gas. 
This is a typical assumption for clumpy PDR models (see e.g., \citealt{2010Wolfire} and \citealt{2014ApJ...784...80L}). 
In this scenario, a measured value of $A_V$ through the
cloud could be made up of clumps of size less than $A_V$ or
a single clump of size $A_V$, both of which look the
same to the model where we integrate continuously up
to $A_V$.

\subsection{Comparison with [CII] Observations}

For each position, the PDR model predicts the integrated [CII] and CO intensities. We plot these model predictions and observations for branch A in Figure ~\ref{f:A_Theory}.
The PDR model predicts the flat trend of $I_{\rm [CII]}$ to within 2$\sigma$ and the general decreasing trend of $I_{\rm CO}$ as a function of decreasing $A_V$. 
This demonstrates that a steady-state chemical equilibrium model with a core+halo density structure, where the halo is more extended spatially than the core, can explain well the observed trends of $I_{\rm [CII]}$ and I$_{\rm CO}$.
As seen in Figure~\ref{f:B_Theory}, the PDR model predicts similar trends as those seen for branch A for both the $I_{\rm [CII]}$ and the I$_{\rm CO}$ for branch B.
The model predictions for branch B appear consistently higher, but within 2$\sigma$, than the observations even with a factor of 2 lower ISRF than was used for the model of branch A positions.
While CO trends have been reproduced well by the model, for both branches there is a slight disagreement
at the points with the highest $A_V$. 
This may be caused by additional heating from a central source such as
FUV photons or cosmic-rays \citep{2019ApJ...878..105G, 2019ApJ...883..190G}
that are not accounted for by external heating alone.


Overall, the observations and the predictions both show a flat profile of [CII] emission for both branches A and B, indicating that the incident radiation field, heating rate, and column density of C$^+$ are relatively constant across the individual branches (see section~\ref{sec:IntIntComp}). 
We note that most of the C$^+$ emission arises in the atomic gas which
is relatively constant between branches. The only difference between
branches A and B is the need for about 2 times lower ISRF in branch
B relative to branch A. Keeping the ISRF field constant between
branches would overproduce the C$^+$ intensity in branch B arising from the atomic
gas alone.
However, the model does not predict a significant ``CO-dark" molecular gas component in either branch A or branch B, due to the ad-hoc density distribution which abruptly jumps from
diffuse to dense gas. The same model was used in \citet{2014ApJ...784...80L} and successfully
explained the observed CO  and $X_{\rm CO}$ factor trends with $A_V$.


The success of the core+halo PDR model without the ``CO-dark'' molecular gas in explaining the observed [CII] profiles suggests that Perseus may not have a significant ``CO-dark'' H$_2$ component.
While this result would contradict derivations from L12, it is possible that the IR-derived H$_2$ is affected by the use of a single DGR throughout Perseus. 
A single, instead of varying (due to dust grain evolution) DGR would result in an overestimate of the H$_2$ distribution, and the ``CO-dark'' H$_2$ component.
For example, a DGR of $2 \times 10^{-21}$ mag cm$^2$, which is two times higher than the constant DGR used by L12, would result in essentially no ``CO-dark'' H$_2$ gas on the eastern side of Figure 1. 
While this estimate is illustrative only, it clearly shows that accurate measurements of DGR in and around GMCs are important to constrain the amount of the ``CO-dark'' gas.

It is likely that a more realistic density distribution, perhaps
constant in thermal pressure, or one derived from a turbulent model
might be more successful in predicting the ``CO-dark" molecular gas component.
A similar clumpy model was adopted by L12 and \citet{2010Wolfire}.
However, Wolfire's PDR model includes broadening by turbulence in the form of larger, non-thermal FWHMs, but does not encompass most of the added characteristics of a MHD model.
In the MHD model of \citet{2010Glover} for example, the C$^{+}$ abundance peaks at $A_V = 1$ mag and 
then declines with higher $A_V$ as the abundance of CO increases. 
A strikingly large scatter has been found in the MHD simulations; 
there are many regions of high extinction which have a high C$^{+}$ abundance.
This large scatter in the C$^{+}$ abundance is mainly due to the highly 
inhomogeneous density structure generated by turbulence and suggests that
molecule formation is heavily affected by turbulence. 
In addition, Wolfire's PDR model considers only the CNM, while we know that Perseus has a significant fraction of the WNM as well, and a small fraction of thermally unstable HI (\citealt{2014Stanimirovic}, \citealt{2015ApJ...809..122B}).

\section{Conclusions}
\label{sec:Conclusions_6}

We obtained observations of the 158 $\mu$m [CII] emission for two different regions in the Perseus molecular cloud using {\it Herschel}, sampling each region with 20 positions.
Previously, L12 used IR observations to map out the distribution of the ``CO-dark'' H$_2$ across Perseus.
Our branch A samples a region where the IR-derived H$_2$ suggests a significant amount of ``CO-dark'' H$_2$, while branch B probes a region without likely ``CO-dark'' H$_2$.
While studying the spatial extent and properties of the [CII] emission in these two regions, we are also in the position to at least qualitatively compare ``CO-dark'' H$_2$ gas estimates using two independent methods.

In branch A we detected significant $I_{\rm [CII]}$ in almost all 20 positions,
while in branch B, which samples a more diffuse environment, the [CII] emission
is found in only 60\% of positions.
The distributions of $I_{\rm [CII]}$ across each individual branch are relatively flat, with 
the [CII] emission being about two times fainter in branch B.

The observed [CII] emission is extended, reaching $>82'$ ($\sim7$ pc) away from the Perseus center in both branches. 
This is different from the Taurus molecular cloud where \citet{2014Orr} did not detect any significant [CII] emission using {\it Herschel} observations of similar sensitivity. 
The [CII] emission in Perseus is more spatially extended than the $^{12}$CO(1-0) emission. 
The lack of velocity shifts between [CII] emission and the molecular line emission components suggests that there are no clear motions of the transition layer relative to the cloud center.



As the 158 $\mu$m transition is a key coolant for gas with density $<3000$ cm$^{-3}$,
the observed flat $I_{\rm [CII]}$ profiles suggest a relatively uniform heating rate, and a 
uniform incident radiation field, across the two boundary regions, although they probe a significant range of $A_V$, from $\sim1$ to $\sim10$ mag. The observed difference between branches A and B suggests a factor of two higher heating rate, and radiation field, in branch A relative to branch B.

We compared our $I_{\rm [CII]}$, $N({\rm HI})$, and the IR-derived $N({\rm H_2})$ trends with those predicted by the SILCC-Zoom Project simulations of individual GMCs \citep{2018Franeck}. 
We find a good agreement between the HI and total hydrogen (as a proxy of total gas) column densities and $I_{\rm [CII]}$. 
This suggests that the HI envelope plays an important role in explaining the [CII] intensity. 
While Perseus has largely reached chemical equilibrium, it still contains a large HI envelope which
dominates its mass budget and also lacks massive star formation, therefore Perseus appears relatively similar to young GMCs before the onset of massive star formation. 
Comparing the two branches, branch B appears less evolved than branch A. 
The IR-derived H$_2$ column density is higher than that predicted by \citet{2018Franeck} simulations. This could be due to either the simulations underestimating the H$_2$ column density, or 
the IR-derived H$_2$ column density being
overestimated.

Finally, we compared the observed flat $I_{\rm [CII]}$ profiles  with predictions from a 1-D, two-sided slab PDR model \citep{2010Wolfire}. 
The model has a dense core and an extended pure-HI envelope and is tailored specifically for Perseus.
The model accurately predicts the flat $I_{\rm [CII]}$ trends seen in the observations, as well as the trend of decreasing $I_{\rm CO}$ as a function of $A_V$, when an incident radiation field of 0.5 Draine (0.85 Habing) fields is used on each side of the PDR layer for the A branch. 
A factor of two lower radiation field is needed to reproduce $I_{\rm [CII]}$ for branch B.

However, the PDR model has an artificial step-function density
distribution and as a result does not contain any ``CO-dark'' molecular gas.
While this comparison suggests that no ``CO-dark'' molecular gas is needed to explain the observed $I_{\rm [CII]}$ profiles, implementation of a more realistic density structure, that includes a more gradual H$_2$ density distribution and ``CO-dark" H$_2$, would likely produce equally accurate predictions.
However, a factor of two higher DGR on one side of Perseus implemented into the IR derivation of H$_2$ would result in the absence of any IR-derived H$_2$.
While we think this is not a very likely scenario, spatial variations of DGR across Perseus require further investigations. 


In summary, 
\begin{enumerate}
    \item The highly extended ($\sim$7 pc) [CII] emission in Perseus is associated predominantly with the CNM from the Perseus HI envelope. 
    \item A steady-state chemistry, PDR model successfully reproduced a relatively uniform integrated intensity of [CII] emission, while employing a step-function density distribution with a highly extended, $>$3$\times$ the size of the core, pure HI envelope and dense core of HI and H$_2$. 
    \item The difference in [CII] intensity between branches A and B can be explained by a factor of two difference in the incident radiation field. 
    \item At the first level, assuming a constant DGR, there is no need to invoke ``CO-dark" H$_2$ gas to explain observed properties of the [CII] emission in Perseus.
    \item ``CO-dark'' H$_2$ gas calculations require detailed considerations of the density distribution in the PDR model, as well as further constraints on the spatial variations of the DGR across Perseus. 
\end{enumerate}

\acknowledgements
This work is based [in part] on observations made with {\it Herschel}, a European Space Agency Cornerstone Mission with significant participation by NASA. Support for this work was provided by NASA through an award issued by JPL/Caltech.

S.S. acknowledges the support provided by the NSF Early Career Development (CAREER) Award AST-1056780, the Vilas funding provided by the University of Wisconsin, and the John Simon Guggenheim fellowship. M.-Y.L. was partially funded through the sub-project A6 of the Collaborative Research Council 956, funded by the Deutsche Forschungsgemeinschaft (DFG). 
The authors are grateful to Jay Gallagher for stimulating discussions, and an anonymous referee for constructive suggestions. This research was carried out in part at the Jet Propulsion Laboratory, which is operated for NASA by the California Institute for Technology. This research made use of Astropy, a community-developed core Python package for Astronomy 37. 

\software{Astropy \citep{astropy:13,astropy:18}, MIRIAD \citep{Sault95}, KARMA \citep{Gooch95}, SciPy \citep{Oliphant07, scipy_2}, NumPy \citep{numpy, numpy2}, Matplotlib \citep{Hunter07}}

\facilities{Herschel, Arecibo, FCRAO, CTIO:2MASS}

This work is based [in
part] on observations made with the {\it Spitzer Space Telescope}, which is operated by the Jet Propulsion
Laboratory, California Institute of Technology under a contract with NASA.

The {\it Herschel Space Telescope} is an ESA space observatory with science instruments provided by European led 
Principal Investigator consortia and with important participation from NASA.
HIFI has been designed and built by a consortium of institutes and university departments from across Europe, Canada and the United States under the leadership of SRON Netherlands Institute for Space Research, Groningen, 
The Netherlands and with major contributions from Germany, France and the US. Consortium members are: Canada: CSA, U.Waterloo; France: CESR, LAB, LERMA, IRAM; Germany: KOSMA, MPIfR, MPS; Ireland, NUI Maynooth; Italy: ASI, IFSI-INAF, Osservatorio Astrofisico di Arcetri-INAF;
Netherlands: SRON, TUD; Poland: CAMK, CBK; Spain: Observatorio Astronómico Nacional (IGN), Centro de Astrobiología (CSIC-INTA). Sweden: Chalmers University of Technology - MC2, RSS \& GARD; 
Onsala Space Observatory; Swedish National Space Board, Stockholm University - Stockholm Observatory; Switzerland: ETH Zurich, FHNW; USA: Caltech, JPL, NHSC.

The Arecibo Observatory is operated by SRI International under a cooperative agreement with the National Science Foundation (AST-1100968), and in alliance with Ana G. Méndez-Universidad Metropolitana, and the Universities Space Research Association.

\appendix
  \label{sec:App}





\section{{\it Herschel} Data Reduction}
\label{s:app-reduction}

Many off-source, reference spectra were contaminated with emission centered on or near the central [CII] peak at a velocity of 4-9 km\hspace{0.5pt}s$^{-1}$. 
This led to the data-reduction process without reference spectra subtraction as outlined in Section \ref{sec:Obs_Data_3}.

\begin{figure*}[htp]
 \includegraphics[width=\textwidth]{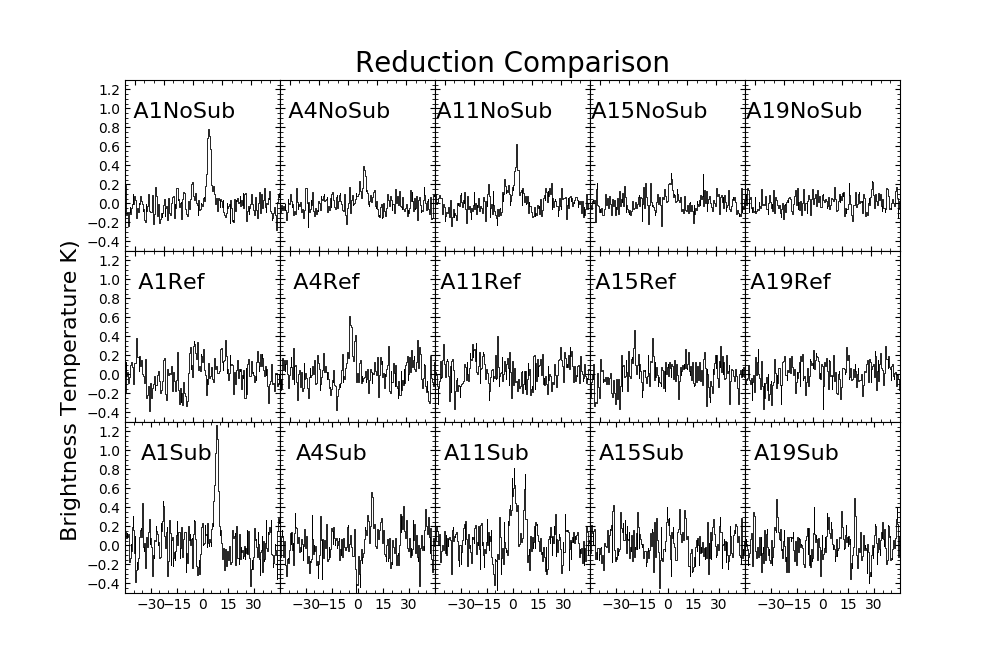}
  \caption{Top row: A11-A15 spectra, using the reduction method without reference spectra subtraction. Middle Row: the reference spectra for each position. Bottom Row: A11-A15 with reference spectra subtraction.}
  \label{f:MethodCompare}
\end{figure*}

To verify the accuracy of our reduction method, the reduction method using the reference spectra subtraction was applied for spectra without obvious contamination (e.g. Figure ~\ref{f:MethodCompare} second row, ``A1ref", the reference spectrum for position A1). 
The resultant spectra are then compared with the results from our reduction method, as can be seen in Figure~\ref{f:MethodCompare}.
In Figure~\ref{f:MethodCompare} the top row shows spectra produced with our reduction method (without using the reference spectrum), the middle row shows the reference spectra, and the bottom row shows spectra with reference spectra subtracted. 
The specific positions were chosen to show a diversity of strong and weak emission and with obvious reference spectrum contamination or no reference spectrum contamination.

For positions where there is no obvious reference spectrum contamination, e.g. A11, the spectral profiles produced by two methods look reasonably similar and their peaks are in agreement within 2-$\sigma$, showing that the method without reference spectrum subtraction is reasonable and is not yielding false detection. 
The method using the reference position has a $\sqrt{2}$ higher noise level, with
$\sigma=0.085$ K as opposed to $\sigma=0.06$ K for the method without reference subtraction.  
In some cases, e.g. A15, the peak of emission within the no reference spectrum subtraction spectrum is just below the noise level and this is why the spectral line appears in the no-reference method while it is buried within the noise in the reference-method spectrum. In some cases, e.g. A1, the no-reference spectrum may be slightly underestimating the peak intensity.
To have a uniformly processed data set, all spectra were processed using the method without reference spectra subtraction.

As a second check of our reduction method, we applied our methodology to 158 $\mu$m data from \citet{2014Orr} for the PDR region in the Taurus molecular cloud. 
Our method yielded similar noise level for spectra with $\sigma= 0.032$ K vs the published average value of 0.04 K. If instead we use the reference positions, as did \citeauthor{2014Orr},  we get $\sigma= 0.048$ K for the same test spectrum, which is a factor of $\sqrt{2}$ higher than when not using the  reference subtraction.
This check further indicates that our reduction method is comparable to the standard method and is not introducing artifical spectral lines.

\begin{figure*}[htp]
 \includegraphics[width=\textwidth]{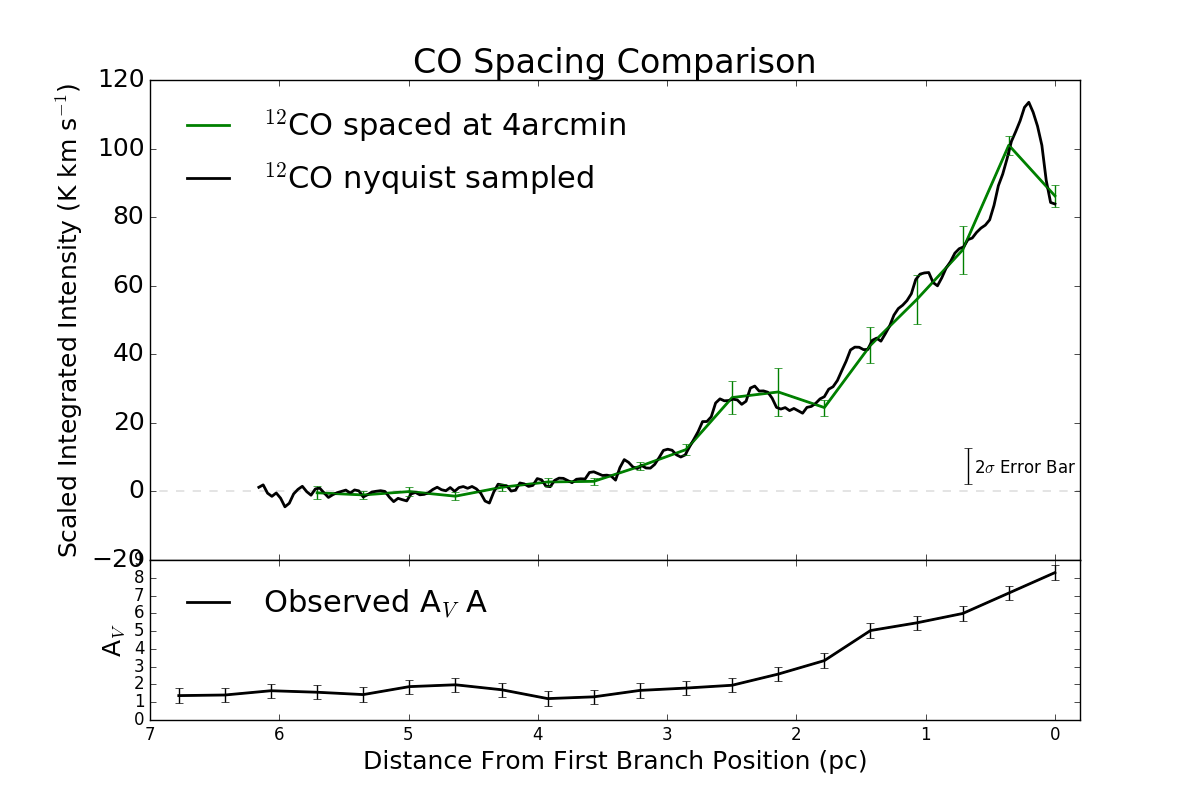}
  \caption{A comparison of COMPLETE $^{12}$CO data sampled at every 4.1' in green, as opposed to Nyquist sampling the same data in black \citep{2006Ridge}. COMPLETE $^{12}$CO data has a pixel size of 23$\arcsec$ which has been regridded from the 46$\arcsec$ beam FWHM of the FCRAO telescope at 115GHz. The bottom plot shows the $A_V$ as a function of position, it has a resolution of $\sim$4.1$\arcmin$. The representative 2$\sigma$ error bar for the Nyquist sampled data is shown in the bottom right of the top plot.}
  \label{f:CO_Comp}
\end{figure*}

\section{Beam Dilution}
\label{s:app-beam} 
 
A telescope beam can usually be approximated with a Gaussian function of size $\Theta_s$, and if we assume that the source has a Gaussian shape, then the actual or true source brightness temperature $T$ is related to the main-beam brightness temperature observed by the telescope $T_{mb}$ by:
\begin{equation}
T= T_{mb} \frac{\theta_s^2 + \theta_{beam}^2}{\theta_s^2}    
\end{equation}
where $\theta_s$ is source's true angular size and $\theta_{beam}$ is the telescope beam size (\citealt{2018tra..book.....W}, eq. 7.22). 
In the analyses in this paper we use several datasets with different angular resolution. 
This means that when comparing two datasets at different resolution, their brightness temperature will be affected (based on the above equation), or diluted, by a different amount. 
To estimate whether our results are affected by beam dilution, we follow the method from \citet{2017Pineda}. 
This method provides only a first-order correction of beam dilution as usually the ISM structure is more complex than a simple Gaussian representation.


If we are observing a single source at the same frequency with two different telescopes, where $\theta_a$ and $\theta_b$ are two different telescope beam sizes, then the dilution factor can be expressed by (based on equation (14) and (15) from \citealt{2017Pineda}): 
\begin{equation}
f_{1,2}= \frac{ \theta_s^2 + \theta_a^2 }{\theta_s^2 + \theta_b^2} = \frac{T_{mb,2}}{T_{mb,1}},
\end{equation}
where $T_{mb,1}$ and $T_{mb,2}$ are the peak brightness temperatures of the source when observed at two resolutions.
Essentially, the ratio of peak main-beam brightness temperature at two different resolutions can be used to  estimate the dilution factor. 


We compare the {\it Herschel} dataset with angular resolution of 12$\arcsec$ with FCRAO's $^{12}$CO data having a resolution of 46$\arcsec$ and the HI and H$_2$ datasets with a resolution of 4$\arcmin$. 
To assess if our comparisons would have the same conclusions if the [CII] observations had the same resolution of that of CO or HI, we perform the following tests. 
First, we use the 
{\it Spitzer} 8 $\mu$m image of Perseus \citep{2003PASP..115..965E, 2007Evans} to assess clumpiness of the [CII] emission.  The 8 $\mu$m emission largely traces PAHs which provide the key heating source (via photoelectric effect) in the diffuse ISM \citep{1997ARA&A..35..179H}. In thermal equilibrium, heating and cooling (largely via [CII] emission) balance out, and the 8 $\mu$m emission can be used to gauge small-scale structure of the [CII] emission.
This strategy is based on the assumption that the 158 $\mu$m line is the dominant coolant in the neutral ISM. 
Another major coolant of PDRs is the [OI] 63 $\mu$m transition, but given the incident ISRF, the two PDRs we are studying in Perseus do not fall within the temperature regime where [OI] emission contributes much to the overall cooling.
Previous studies of Perseus suggest the volume density at $<$100 cm$^{-3}$ for all but the densest portions of Perseus. 
For a FUV ISRF of $\sim$1G$_0$ and a density of 100 cm$^{-3}$ the [OI]/[CII] ratio is expected to be $<0.1$ \citep{2007ASPC..375...43K}.
The Perseus extended envelope has a volume density closer to 40 cm$^{-3}$ (from L12), which pushes the expected [OI]/[CII] ratio closer to 0.03 \citep{2007ASPC..375...43K}.
Given these expected [OI]/[CII] ratios, we conclude that the [OI] emission is not a significant fraction of the cooling budget within the two boundary regions in this study and the [CII] emission can be assumed to trace the cooling and can be reflected by the PAH population and its emission.

\begin{figure*}[htp]
 \includegraphics[width=\textwidth]{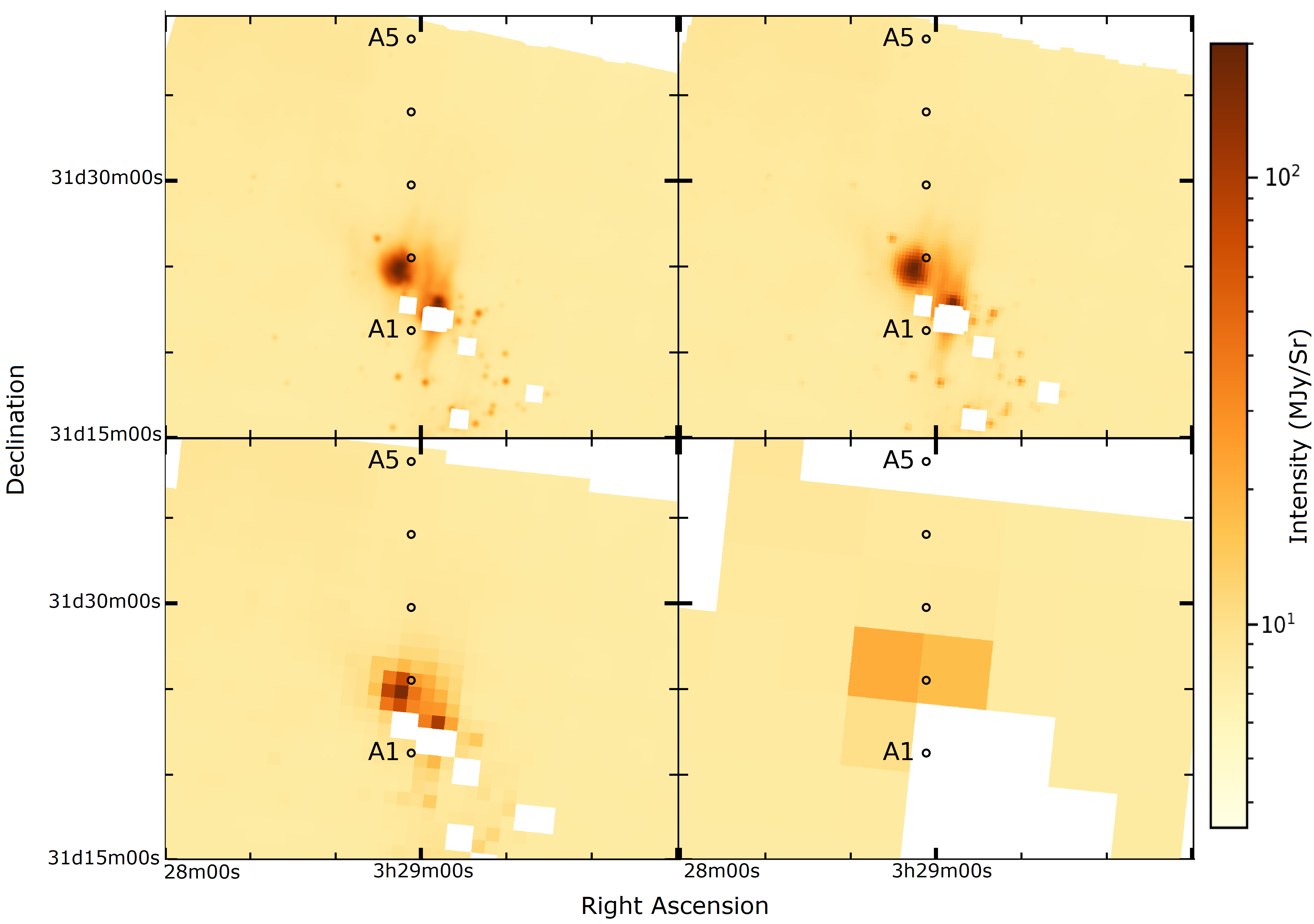}
  \caption{The {\it Spitzer} IRAC intensity image at 8 $\mu$m shown at different resolutions. The intensity is in MJy/Sr and displayed in log-scale.
  All subplots have the first 5 positions of branch A over-plotted as circles. Each position's circle is scaled to {\it Herschel}'s beam size at 158 $\mu$m, 12$\arcsec$. 
  White pixels are oversaturated due to coinciding
  with stellar positions and have Nan values.
  When convolving the image, we did not extrapolate over these pixels, any area with a NaN value in it returned a NaN for the whole area's new value. Top left: The image at its native resolution of 1.2$\arcsec$. Top right: The image convolved to the resolution of the {\it Herschel} [CII] observations, 12$\arcsec$. Bottom left: The image convolved to the resolution of the $^{12}$CO image, 46$\arcsec$. Bottom right: The image convolved to the resolution of the HI dataset, 4$\arcmin$.}
  \label{f:diff_res}
\end{figure*}

The {\it Spitzer} 8 $\mu$m image of Perseus has a resolution of  1.2$\arcsec$.
We convolve the 8 $\mu$m  emission, centered at pointings A1-A5, to 12$\arcsec$, 46$\arcsec$ and 4$\arcmin$ by averaging the 8 $\mu$m intensity within appropriate circular apertures. 
Figure ~\ref{f:diff_res} shows the original image at 1.2$\arcsec$, as well as the image once convolved to a resolution of 12$\arcsec$, 46$\arcsec$, and 4$\arcmin$.
Positions A1-A5 are displayed on the plots for easy reference.
Positions A6-A20 and all of the B branch are not included in this analysis due to lack of {\it Spitzer} 8$\mu$m coverage.
Figures 2 and 3 show that the [CII] emission is fainter and more diffuse as we go further away from the base of branch A. Since positions A1-A5 are not significantly effected by the beam dilution, we expect that the positions (A6-A20), which probe a more diffuse medium and are further from NGC1333, are even less so. The same logic applies to branch B, which is far from known massive stars and is even more diffuse than positions A1-A5 as indicated by the A$_V$ map.



The results are shown in  Table \ref{tab:table_spitz_append}. 
For each pointing, within uncertainties, the intensity per pixel does not change much when we smooth the 8 $\mu$m emission to the resolution of
12$\arcsec$, 46$\arcsec$ and 4$\arcmin$. This suggests that the corresponding dilution factors are close to unity, which means that the 8 $\mu$m and the [CII] emission have predominantly smooth, diffuse distributions. Therefore, the results in the paper are not significantly affected by different resolutions. 
This conclusion is also supported by the comparison between the [CII] intensity for the A2 pointing with previous observations by  \citet{2002Young} which had lower resolution than the {\it Herschel} observations yet observed intensities agree very well. Again, this demonstrates that the [CII] emission in Perseus largely comes from diffuse gas.

To investigate how the visual extinction was affected by beam dilution we apply the same procedure to the high resolution column density image created by \citealt{2013A&A...550A..38P} using the {\it Herschel} 70 $\mu$m, 160 $\mu$m PACS data as well as the 250 $\mu$m, 350 $\mu$m, 500 $\mu$m SPIRE images from the {\it Herschel} Gould Belt Survey \citep{2010A&A...518L.102A, 2012A&A...547A..54P}.
We use this dataset because it has high resolution (18.2$\arcsec$), and the total (dust) column density correlates with the observed visual extinction (e.g. \citealt{2010A&A...518L..88B}).
After this we undertook the same analysis as mentioned above for the 8 $\mu$m and show our results for all pointings of the A branch in Table \ref{tab:table_dust_append}. Similarly to the 8 $\mu$m analysis, 
we find that the total column density (and correspondingly A$_V$) does not suffer from severe beam dilution.
Data for branch B do not exist, but be assume that based on other datasets, branch B suffers from beam dilution on the same scale as branch A, if not less so due to it being more diffuse.

In addition to the above analysis we also note that
the $^{12}$CO dataset (from the COMPLETE survey, \citealt{2006Ridge}) has a finer angular resolution than the sampling scale of 4$\arcmin$ (driven by the resolution of HI and H$_2$ datasets). We therefore check whether the coarse sampling could be affected by small-scale fluctuations of the $^{12}$CO emission.
To do this, we compared the $^{12}$CO profiles sampled at 4$\arcmin$ with a profile sampled at the Nyquist rate of 46$\arcsec/2=23\arcsec$.
We have done this for the branch A and results are shown in Figure~\ref{f:CO_Comp}.
The two profiles agree very well and we conclude that sampling at 4$\arcmin$ is not missing any
important structure in the $^{12}$CO distribution.




\begin{table*}[htp]
\centering
\begin{tabular}{|l|c|c|c|}
\toprule
 Position & 12$\arcsec$ & 46$\arcsec$ & 4.0$\arcmin$ \\
\hline
 A1  &  $9.6 \pm 0.3$  & $9.6 \pm 0.3$ & NaN \\
\hline
 A2  &  $32.9 \pm 0.2$  & $34.0 \pm 0.2$ & $35.7 \pm 0.5$ \\
\hline
 A3  &  $9.2 \pm 0.2$  & $9.2 \pm 0.2$ & $9.1 \pm 0.2$ \\
\hline
 A4  &  $8.8 \pm 0.2$  & $8.8 \pm 0.2$ & $8.8 \pm 0.2$ \\
\hline
 A5  &  $8.9 \pm 0.2$  & $8.9 \pm 0.2$ & NaN \\
\hline
\end{tabular}
\caption{Table of average MJy/Sr per pixel of the {\it Spitzer} 8$\mu$m intensity image data \citep{2003PASP..115..965E, 2007Evans}. The average per pixel refers to the average per smallest area resolution (2$\arcsec$ by 2$\arcsec$), but averaged over a circular aperture with a diameter the same as the angular resolution of {\it Herschel} at 157.7 $\mu$m: 12$\arcsec$, FCRAO at 115 GHz: 46$\arcsec$, and of Arecibo at 21 cm: 4$\arcmin$.}
\label{tab:table_spitz_append}
\end{table*}

\begin{table*}[htp]
\vspace{-20.0pt}
\centering
\begin{tabular}{|l|c|c|c|}
\toprule
Position &            12$\arcsec$ &               46$\arcsec$ &               4.0$\arcmin$ \\
\hline
      A1 &    248.5$\pm$9.5 &  294.8$\pm$89.8 &  233.7$\pm$204.7 \\
      \hline
      A2 &     41.8$\pm$0.7 &   41.2$\pm$2.1 &   54.2$\pm$10.6 \\
      \hline
      A3 &     43.3$\pm$7.8 &   47.9$\pm$12.0 &   47.2$\pm$15.1 \\
      \hline
      A4 &     20.8$\pm$0.6 &   21.0$\pm$1.3 &   22.2$\pm$2.3 \\
      \hline
      A5 &     16.5$\pm$0.4 &   16.5$\pm$0.7 &   14.8$\pm$1.4 \\
      \hline
      A6 &     11.4$\pm$0.5 &   11.1$\pm$0.8 &   11.6$\pm$0.9 \\
      \hline
      A7 &     10.7$\pm$0.3 &   11.1$\pm$0.4 &   12.0$\pm$0.6 \\
      \hline
      A8 &     15.6$\pm$0.9 &   15.4$\pm$1.7 &   12.7$\pm$2.3 \\
      \hline
      A9 &      7.8$\pm$0.3 &    7.9$\pm$0.5 &    9.0$\pm$0.8 \\
      \hline
     A10 &     11.1$\pm$0.4 &   11.4$\pm$0.9 &    9.6$\pm$1.2 \\
     \hline
     A11 &      6.1$\pm$0.3 &    6.6$\pm$0.3 &    6.7$\pm$0.7 \\
     \hline
     A12 &      6.7$\pm$0.2 &    6.7$\pm$0.4 &    6.6$\pm$0.5 \\
     \hline
     A13 &      7.0$\pm$0.3 &    6.9$\pm$0.5 &    6.7$\pm$0.5 \\
     \hline
     A14 &      5.9$\pm$0.2 &    6.1$\pm$0.4 &    6.3$\pm$0.5 \\
     \hline
     A15 &      6.3$\pm$0.4 &    6.3$\pm$0.3 &    6.4$\pm$0.5 \\
     \hline
     A16 &      5.4$\pm$0.4 &    5.8$\pm$0.5 &    6.2$\pm$0.5 \\
     \hline
     A17 &      6.3$\pm$0.4 &    6.2$\pm$0.6 &    6.3$\pm$0.6 \\
     \hline
     A18 &      6.0$\pm$0.2 &    6.1$\pm$0.5 &    6.2$\pm$0.6 \\
     \hline
     A19 &      nan &    nan &    nan \\
     \hline
     A20 &      nan &    nan &    nan \\
\hline
\end{tabular}
 \caption{Table of average total column density in units of 10$^{20}$ cm$^{-2}$ per pixel for the Gould Belt Survey high-resolution column density image \citep{2010A&A...518L.102A, 2012A&A...547A..54P, 2013A&A...550A..38P} with errors. The average per pixel refers to the average per smallest area resolution (18.2$\arcsec$ by 18.2$\arcsec$), but averaged over a circular aperture with a diameter the same as the angular resolution of {\it Herschel} at 157.7 $\mu$m: 12$\arcsec$ (1 pixel diameter), FCRAO at 115 GHz: 46$\arcsec$ (2.5 pixel diameter), and of Arecibo at 21 cm: 4$\arcmin$ (13.2 pixel diameter). The errors are the standard deviation of the column density within each aperture.}
\label{tab:table_dust_append}
\end{table*}

\end{document}